\documentclass[final]{IEEEtran}
\IEEEoverridecommandlockouts
\usepackage{cite}
\usepackage{amsmath,amssymb,amsfonts,amsthm}
\usepackage{bbm}
\usepackage[most]{tcolorbox}
\usepackage{graphicx}
\usepackage{multirow}
\usepackage{textcomp}
\usepackage{pifont}    
\usepackage{xcolor}
\usepackage{cite}
\usepackage{enumitem}
\usepackage{algorithm}
\usepackage{algpseudocode}
\usepackage{subcaption}
\usepackage[normalem]{ulem}

\def\BibTeX{{\rm B\kern-.05em{\sc i\kern-.025em b}\kern-.08em
		T\kern-.1667em\lower.7ex\hbox{E}\kern-.125emX}}

\def\nb0{{\mathbf{0}}}
\def\nb1{{\mathbf{1}}}

\newtheorem{definition}{Definition}

\newtheorem{remark}{Remark}

\def\figref#1{Fig.\,\ref{#1}}%
\def\eqnref#1{Eqn.\,\ref{#1}}%

\begin{document}
	\newcommand{\cmark}{\ding{51}} 
	\newcommand{\xmark}{\ding{55}} 
	\title{Full Motion State Localization with Extra
		Large Aperture Arrays}
	
	\author{Wasif J. Hussain, Don-Roberts Emenonye, R. Michael Buehrer, Harpreet S. Dhillon \,\,\vspace{-2.1em}
		\thanks{The authors are with Wireless@VT, Virginia Tech, Blacksburg,	VA 24061 USA (e-mail: {wasif, donroberts, rbuehrer, hdhillon}@vt.edu). The support of US NSF (grant CNS-2107276) is gratefully acknowledged.}
	}
	\maketitle
    
	\begin{abstract}
        Conventional localization techniques typically assume far-field (FF) propagation characterized by planar wavefronts and simplified spatial relationships. The use of higher carrier frequencies has given rise to the paradigm of extra large aperture arrays (ELAAs) which consist of a large number of tightly packed antenna elements. These arrays have a large electrical aperture which increases the Fraunhofer distance making the FF assumption restrictive. As a result, near-field (NF) effects, such as spherical wavefront curvature, direction dependent gains, and spatial variations in Doppler and delay, become significant even at distances previously regarded as FF. This paradigm shift opens up new opportunities: the rich multi-parametric structure of NF models if properly exploited can enable superior localization accuracy. In this work, we investigate the potential of multi-snapshot, full-motion state ($3$D position, $3$D velocity, and $2$D orientation) estimation using delay and Doppler measurements for a mobile receiver equipped with a linear ELAA in an environment comprising a number of wideband anchors. We develop a signal model that captures both the NF propagation geometry and spatially varying Doppler effects. We perform an information-theoretic analysis to establish Cram\'er--Rao lower bounds (CRLB) on the achievable position error bound (PEB), velocity error bound (VEB), and orientation error bound (OEB), respectively. Using these results, we determine the minimum infrastructure required to achieve full-motion state localization, and extend the effort on NF sensing based on Doppler measurements. We reveal that delay measurements carry richer information than Doppler measurements, and standalone Doppler measurements cannot overcome information losses due to unknown channel gains and frequency offsets, enabling only coarse estimation capabilities. We also propose a maximum-likelihood (ML) approach to jointly estimate the $8$D position parameters from measured channel characteristics. 
	\end{abstract}
	
	
	\section{Introduction}
	
	Wireless systems are built, optimized, and deployed under the assumption that the transmitter and receiver operate in the far-field (FF) propagation regime. However, the use of higher carrier frequencies and the proposal to implement extra large aperture arrays (ELAA) consisting of hundreds of thousands of densely packed antenna elements in apertures spanning several meters have challenged this assumption \cite{ye2024extremely}. For clarity, the region closest to the transmitter is always a near-field (NF) region. When a receiver is located in this region, it observes substantial wavefront curvature, which is spherical. As the receiver moves farther away and enters the FF region, the received wavefront can be accurately approximated as planar. The boundary separating these two propagation regions is called the Fraunhofer boundary, and it is proportional to the aperture diameter and the operating frequency. An ELAA system extends the NF propagation regime to several hundreds of meters, and substantial work has investigated how additional information in the spherical wavefront can improve beamforming \cite{Zhang2022_NearFieldBeamFocusing}, correct the orientation of reconfigurable intelligent surfaces (RIS) \cite{Emenonye2023_RISMisorientation}, enhance RIS aided positioning \cite{Emenonye2023_RISLocOffsets}, and enable target tracking in the near field \cite{Guerra2021_NearFieldTracking}.
	
	Initial investigations of the NF have been communication centric, focusing on improved beamforming and increased data rates. However, the presence of a spherical rather than planar wavefront also motivates the study of NF effects on localization systems \cite{Emenonye2024_SingleAnchorLimits,handbook_of_loc}. The basic intuition is that a spherical wavefront illuminates the receiver such that each antenna element behaves like a virtual anchor, reducing the number of physical anchors or measurements needed for localization. Although NF effects can improve localization accuracy, the fundamental limits of such systems must first be established and understood for three dimensional positioning as well as for velocity and orientation estimation. Hence, in this paper, we use both Fisher information and constrained estimation theory to investigate the feasibility and construct a practical estimator, respectively,  for achieving full motion state localization of a receiver equipped with an ELAA which is receiving downlink signals from wideband anchors while being in their NF. We examine the impact of delay and Doppler in the NF, as well as the effects of the number of anchors, time slots, receiver antennas, and their interactions with time and frequency offsets.

	\begin{table*}[t]
		\centering
		\vrule
		\begin{tabular}{ |c||c|c|c|c|c|c|c|c|c| } 
			\hline
			References &FF/NF&Doppler measurements& Position & Velocity &Orientation& Synchronization error &Information analysis &Estimator design \\
			\hline
			\hline

            \cite{Emenonye2023_RISLocOffsets} & Both & \xmark & \cmark &\xmark &\cmark &\cmark & \cmark &\xmark \\ 
			\hline
            
			\cite{garcia2017direct}& FF&\xmark&\cmark&\xmark&\xmark&\xmark&\xmark &\cmark \\ \hline
			
			\cite{shahmansoori2017position}&FF& \xmark &\cmark&\xmark&\cmark&\xmark&\cmark&\cmark \\ 
			\hline
			
			\cite{mendrzik2018harnessing} & FF&\xmark&\cmark&\xmark&\cmark&\xmark&\cmark&\xmark  \\ 
			\hline
			
			\cite{zeng2021massive} & FF&\xmark&\cmark&\cmark&\xmark&\xmark&\xmark&\cmark  \\ 
			\hline
            
			\cite{8804387} & FF&\cmark&\cmark&\cmark&\cmark&\cmark&\cmark&\xmark\\
			\hline
			
			\cite{gong2023simultaneous} & FF & \cmark &\cmark&\xmark&\xmark&\xmark&\cmark&\cmark \\
			\hline

            \cite{chen2024elaa} & NF & \xmark & \cmark & \xmark & \xmark &\xmark &\cmark &\cmark\\
			\hline
			
			\cite{guerra2018single} & NF & \xmark & \cmark &\xmark &\cmark &\cmark & \cmark & \xmark \\
			\hline
			
			\cite{guerra2021near} & NF & \xmark & \cmark & \cmark & \xmark & \cmark & \cmark & \cmark \\
			\hline
			
			\cite{jiang2024near} & NF & \xmark &\cmark & \cmark & \xmark & \xmark & \xmark & \cmark \\
			\hline
			
			\cite{yuan2024scalable} & NF & \xmark & \cmark & \xmark & \xmark & \xmark & \cmark & \cmark \\
			\hline

            \cite{wei2025fundamental} & NF & \cmark & \cmark & \cmark & \xmark & \xmark & \cmark & \xmark \\
			\hline

            \cite{wei2025fundamental2} & NF & \cmark & \cmark & \cmark & \xmark & \xmark & \cmark & \xmark \\
			\hline
            
            \cite{liu2024low} & NF & \xmark & \cmark &\xmark &\xmark &\xmark &\xmark &\cmark \\
			\hline
			
			\cite{wang2025performance} & NF & \xmark & \cmark &\xmark &\xmark &\xmark &\cmark &\cmark \\
			\hline
			
			\cite{hu2018beyond} & FF & \xmark & \cmark & \xmark &\xmark &\xmark &\cmark &\xmark \\
			\hline
			\cite{emenonye2023fundamentals} & FF & \xmark & \cmark & \xmark &\cmark & \xmark &\cmark &\xmark \\
			\hline
			
			\cite{dardari2021nlos} & NF & \xmark & \cmark &\xmark &\xmark & \cmark & \cmark & \cmark \\
			\hline 
			
			\cite{abu2021near, han2022localization,pan2023ris} & NF & \xmark & \cmark & \xmark &\xmark & \xmark & \cmark & \cmark \\
			\hline
			
			\cite{rahal2024ris} & NF & \xmark &\cmark &\cmark &\xmark & \cmark & \cmark & \cmark \\
			\hline
			
			\cite{wang2024near} & NF & \cmark & \xmark & \cmark & \xmark &\xmark &\cmark & \cmark \\
			\hline    
			
			\cite{perf_bound_vel_est_elaa} & NF & \cmark & \xmark & \cmark & \xmark & \xmark & \cmark &\xmark \\
			\hline       
			\cite{emenonye2025joint9dreceiverlocalization} & FF & \cmark & \cmark & \cmark &\cmark & \cmark &\cmark &\xmark \\
			\hline
			
			\cite{meng2025near} & NF & \cmark & \cmark & \cmark & \xmark & \xmark & \cmark & \cmark \\
			\hline
			
			This work&NF & \cmark & \cmark & \cmark & \cmark &\cmark & \cmark &\cmark \\
			\hline
		\end{tabular}
		\caption{Overview of existing research efforts.}
		\label{tab:compar}
	\end{table*}

	\subsection{Prior Art}
	
	The following three research directions are of interest to this paper: 
	(i) localization with large arrays, 
	(ii) RIS aided localization, and 
	(iii) Doppler based velocity sensing in the near field. 
	Relevant prior works from these directions are discussed next. 
	
	
	
	\subsubsection{Localization with large arrays}
	The use of higher-frequency bands has enabled the deployment of a large number of antenna elements, and the resulting ELAAs can be used for communication and sensing. ELAAs have been popularly studied for localization purposes in the context of massive MIMO \cite{garcia2017direct,shahmansoori2017position,mendrzik2018harnessing,zeng2021massive,8804387,gong2023simultaneous, liu2024low, wang2025performance}. The existing massive MIMO architecture for communication, which exploits large antenna apertures for spatial multiplexing and beamforming gains, can be utilized to enable high resolution angular and delay estimation for precise localization. In \cite{garcia2017direct}, the authors presented a hybrid angle of arrival (AoA) and coarse time of arrival (ToA) based $2$D localization approach in a distributed massive MIMO setup. The setup in that work combined two suboptimal approaches: first, channel parameters are estimated, and then position is inferred. In \cite{shahmansoori2017position}, the authors derived performance bounds and proposed an algorithm for $2$D position and $1$D orientation estimation for a synchronized MIMO system. In \cite{mendrzik2018harnessing}, prior synchronization was assumed, and the Fisher information was employed to analyze the role of non light of sight (NLoS) components for a similar problem involving $2$D position and $1$D orientation estimation in a mmWave massive MIMO setup. In \cite{zeng2021massive}, the authors proposed a high accuracy $2$D target localization and $2$D velocity tracking framework by leveraging the statistical focusing effect of the autocorrelation function in massive MIMO systems. In \cite{8804387}, the authors investigated the fundamental limits of a mmWave MIMO-OFDM system for $2$D position, $1$D orientation, and $2$D velocity estimation of a target under synchronization offsets and NLoS conditions. In \cite{gong2023simultaneous}, the authors developed a simultaneous localization and communication (SLAC) framework using massive MIMO and OTFS modulation to address high mobility mmWave scenarios. By incorporating spatial wideband effects and delay--Doppler domain modeling, they derived channel parameters and localization CRLBs and designed low complexity estimators for $3$D positioning using ToA, AoA, and Doppler shifts.  The discussed works so far considered a FF operation. Next we discuss some works which have explicitly considered NF operation. In \cite{chen2024elaa}, the authors analyzed the impact of single base station (BS) 3D localization under blockages and proposed a novel blockage detection algorithm. In \cite{guerra2018single}, CRLB performance bounds are derived for a single anchor 3D position and 2D orientation localization. In massive array regimes it is demonstrated that the effect of multi-path is negligible under certain conditions regardless of the SNR condition and array configuration. In \cite{guerra2021near}, the authors investigate the performance of different Bayesian tracking algorithms for 3D position and 3D velocity estimation using a single large antenna array. Through a CRLB analysis it is demonstrated that loss of positioning information outside the Fresnel region is due to ranging error rather than angular resolution. In \cite{jiang2024near}, the authors developed Extended Kalman filter (EKF), and Adapative gradient descent (AGD) based algorithms for full motion state (2D position and 2D velocity) estimation of a target. A sub-array based approach for 3D localization is presented in \cite{yuan2024scalable}. In \cite{wei2025fundamental,wei2025fundamental2}, fundamental limits for joint estimation of 2D position, 2D velocity and radar cross section (RCS) in an ELAA enabled integrated sensing and communication (ISAC) system was established and analyzed. In \cite{liu2024low}, a back-projection algorithm based NF localization method was developed for XL-MIMO sectored uniform circular arrays (sUCA). An information theoretic analysis for NF sensing with circular arrays along with a performance comparison with linear arrays is presented in \cite{wang2025performance}. 
	It is important to highlight that \cite{8804387,gong2023simultaneous} have revealed Doppler measurements to provide radial velocity information, and \cite{wei2025fundamental,wei2025fundamental2} have analyzed full velocity recovery using ELAAs in a synchronized system. However, the real potential of Doppler measurements for full-motion state parameter estimation in a realistic environment with unknown gains and offsets remains unaddressed. Further, the potential for a full-motion state localization leveraging rich NF operation is not addressed.
    
	
	\subsubsection{RIS aided localization}
	In RIS aided localization, ELAAs are deployed as passive or semi-passive reflective surfaces, facilitating localization by intelligently shaping the wireless propagation environment. RIS enabled ELAAs allow for precise control over the direction and phase of reflected signals, enabling improved coverage and localization in challenging scenarios such as NLoS or blocked environments. In \cite{hu2018beyond}, the authors investigated the potential of using an ELAA based RIS system for positioning and derived the Fisher information matrix (FIM) to analyze achievable $3$D position limits. In \cite{emenonye2023fundamentals}, the authors developed fundamental bounds for the estimation of $3$D position and $2$D orientation in a RIS-aided OFDM system. These works have primarily considered a FF operation. Next we discuss some works considered NF operation. In \cite{dardari2021nlos}, the authors analyzed the potential of a single anchor $3$D localization under NLoS/LoS conditions in a RIS-assisted mmWave OFDM system with time offsets. In that work, the authors developed two positioning algorithms and derived theoretical bounds, assuming the RIS is partially obstructed. In \cite{abu2021near}, the authors studied $3$D localization using a RIS operating in a lens mode with a single receive RF chain. The Fisher information bounds are derived and evaluated for multiple RIS phase profile designs (random, directional, and positional), and a low-complexity three stage estimator based on angle and range search is developed. In \cite{han2022localization}, the authors presented a low overhead joint $2$D localization and channel reconstruction scheme for extra large RISs assisted massive MIMO to identify visibility regions (VR) and reconstruct the channel by exploiting spatially varying NF characteristics. In \cite{pan2023ris}, the authors present a joint channel estimation and $3$D localization approach for an ultra-large RIS in a THz communication system based on the second-order Fresnel approximation of the NF channel model. The approach is based on decoupling and separately estimating the user equipment (UE) distances and AoA. In the same work, a subspace based method and a one-dimensional search are used to estimate the UE's AoAs and distances. In \cite{Emenonye2023_RISLocOffsets}, Fisher information analysis is presented for the $3$D position and $3$D orientation estimation with unknown offsets, and conditions for localizability are established for both FF and NF regimes. The loss of information when the FF model is incorrectly applied to the signals received in NF is also analyzed. In \cite{rahal2024ris}, a single anchor based $3$D position and $3$D velocity estimation is presented for RIS-assisted localization in the event of LoS blockages, considering time and frequency offsets. It is worthwhile noting that Doppler measurement based motion parameter estimation has not been explored. Further, the potential for a full-motion state localization leveraging rich NF operation is not addressed.
	
	\subsubsection{Doppler based NF velocity sensing}
	Hitherto, in the FF, it was known that the Doppler frequency is affected only by the radial velocity component due to the planar nature of wavefront propagation. The concept of Doppler frequency based velocity sensing in the NF was developed in \cite{wang2024near}, where the wavefront curvature in the NF enabled full velocity recovery of the target, \emph{i.e.}, both radial and transverse components. In \cite{perf_bound_vel_est_elaa}, information bounds for velocity sensing with an ELAA were derived, and it was shown that the estimation accuracy of the transverse velocity component decreases with the antenna target distance, unlike the radial component. In \cite{emenonye2025joint9dreceiverlocalization}, the information content in Doppler measurements for full motion-state ($9$D) localization with low Earth orbit (LEO) satellites is analyzed. However, the FF operation is characterized by limited parameter observability due to the lack of spatial variation in the observed Doppler frequency. In \cite{meng2025near}, a sub-array based variational message passing approach is developed for NF joint localization and velocity sensing. Most of the these works employed simplifying assumptions such as knowing LoS vector or considering perfect synchronization. The real potential of Doppler measurements to provide robust information in asynchronized and unknown environment remains unexplored. Further, the potential for a full-motion state localization leveraging Doppler measurements in rich NF operation is not addressed.

	\subsection{Contribution}
	We investigate full motion-state localization using downlink wideband signals transmitted from anchors in the NF of an ELAA equipped receiver. In the NF, the received signal exhibits a spherical wavefront that encodes richer geometric information than the FF planar wavefront, which can be exploited to achieve more precise localization. We quantify this information through a Fisher information based analysis and perform localization using optimization theory. Unlike previous work, this is the first work that rigorously investigates and addresses full motion-state localization potential of ELAAs by explicitly leveraging NF operation, accounting for asynchronization challenges, and developing a practical estimator. A \textit{vis\mbox{-}\`a\mbox{-}vis} comparison between existing research efforts and our contributions is provided in Table~\ref{tab:compar}.
	Our main contributions are listed next.
    
{\em Comprehensive signal model capturing the UWB-NF propagation effects} such as spherical wavefront curvature, element-dependent delay, and spatially varying Doppler frequency, as observed by an ELAA, considering time and frequency asynchronization between anchors and receivers. We derive and explain the NF Doppler formulation showing how the Doppler observed at each antenna element depends on the element specific LoS direction. In contrast to the FF case, where Doppler collapses to the radial velocity component, the NF model provides distinct Doppler observations across the array. This richer structure enables the recovery of all three velocity components, thus extending the scope of NF velocity sensing. 

{\em System level insights via a constrained Cram\'er-Rao bound (C-CRB) analyses}. We derive FIM and perform C-CRB analysis explicitly incorporating information loss from unknown nuisance parameters such as time/frequency offsets and channel gains and enforcing unit-norm constraints on the orientation vector. Through this analysis, we give insights into achievable performance limits, the impact of different system parameters, and the minimum infrastructure required to perform full motion state localization. We reveal that for full motion state localization with single snapshot at least three anchors are required, and with two snapshots the same can be achieved with two anchors. We further reveal that diversity in anchor velocity during successive snapshot can enable single anchor localization with at least four snapshots. However, the same would not be possible if the anchors maintain the same direction of motion over successive snapshots.  Further, we demonstrate that ELAAs enable single snapshot velocity estimation which was previously not possible without NF Doppler measurements. Additionally, we investigate the individual role of Doppler measurements for full-motion state localization and reveal that they lack sufficient information to overcome information losses due to unknown channel gains and frequency offsets. We conclude that to utilize the best from Doppler measurements, another form of richer measurements such as delay or AoA is crucial.

{\em Optimal estimator design for full motion-state recovery}. We design a maximum likelihood (ML) estimator that jointly infers the motion state parameters by exploiting delay and Doppler measurements across all the ELAA elements. The estimator integrates Riemannian optimization for orientation estimation and line-search updates for position, velocity, and synchronization offsets estimation. We also develop a geometric initializer that leverages array geometry and system's rectilinear kinematics to provide reliable starting points. Simulation results confirm that the estimator converges to the C-CRB limits.
		
	\section{System Model}
	We consider $N_B$ wideband mobile anchors with known position and velocities, which will be used to localize a receiver equipped with a uniform linear array (ULA) with $N_U$ antennas. The anchors communicate with the receiver over $N_K$ transmission time slots. Each of the transmission slots spans a duration of $\Delta_t$ seconds. Assuming the array aperture at the anchors is much smaller as compared to the ELAA, we denote the position of anchors during the $k$-th transmission slot as $\mathbf{p}^o_{b,k}$, where $b\in\left\{1,2,\dots,N_B\right\}$ and $k\in\left\{1,2,\dots,N_K\right\}$. The reference antenna element at the receiver, which we take as the first element, without loss of generality, is located at $\mathbf{p}_{U,k}$, where $k\in\left\{1,2,\dots, N_K\right\}$. The positions of anchors and the receiver are defined with respect to a global origin and a global reference axes\cite{handbook_of_loc}. The position of the antenna elements on the receiver during the $k$-th transmission slot is defined with respect to the receiver reference element, indexed by $U$, as $\mathbf{s}_{u,k}=(u-U)d_{\rm a}\bar{\mathbf{s}}_k$, where $d_{\rm a}$ is the antenna spacing, and $\bar{\mathbf{s}}_k$ is a unit vector denoting the ELAA receiver orientation \footnote{A ULA is invariant about rotations around its principal axis. Thus its orientation can be uniquely determined using only two orientation angles, {\em i.e}, pitch and yaw. Therefore, a ULA has only $2$D orientation estimation capabilities. However, with $2$D arrays, the third orientation angle {\em i.e}, roll can be easily estimated, as highlighted in \cite{emenonye20249d,emenonye2025joint9dreceiverlocalization}.} during the $k$-th transmission slot.
	
	We denote the position of the $u$-th receive antenna element during the $k$-th transmission slot with respect to the global origin as $\mathbf{p}_{u,k}=\mathbf{p}_{U,k}+\mathbf{s}_{u,k}$. Let $\mathbf{p}_{bu,k}=\mathbf{p}^o_{b,k}+d_{bu,k}\mathbf{\Delta}_{bu,k}$ denote the position of the $u$-th antenna element with respect to the $b$-th anchor during the $k$-th transmission slot, where $d_{bu,k}$ and $\mathbf{\Delta}_{bu,k}$ are the distance and corresponding unit direction vector from the $b$-th anchor to the $u$-th antenna element of the ELAA during the $k$-th transmission slot, respectively. Let  $\bar{\mathbf{v}}^o_{b,k}$, $\bar{\mathbf{v}}_{u,k}$, and $\bar{\mathbf{v}}_{bu,k}=\bar{\mathbf{v}}^o_{b,k}-\bar{\mathbf{v}}_{u,k}$ denote the velocity of the $b$-th anchor, $u$-th antenna element, and the relative velocity between them during the $k$-th transmission slot, respectively. For the purpose of this work, we assume the receiver and anchors move with constant linear speeds $v_b$ and $v_u$, respectively, across the $N_K$ transmission slots. For practical systems, the attitude dynamics is much slower than translational dynamics and thus we assume the receiver orientation to not change during any of these slots. Therefore, we can write: 
	\begin{equation}
		\begin{split}
			&\mathbf{p}^o_{b,k}=\mathbf{p}^o_{b,0}+(k-1)\Delta_t \bar{\mathbf{v}}^o_{b,k},\\    &\mathbf{p}_{bu,k}=\mathbf{p}_{bu,0}+(k-1)\Delta_t \bar{\mathbf{v}}_{bu,k}. 
		\end{split}
	\end{equation}
	\subsection{Transmit and Receive Processing}
	
	The anchors communicate with the ELAA receiver using quadrature modulation. The $b$-th anchor transmits the following symbol at time $t$ during the $k$-th transmission slot:
	\begin{equation}
		x_{b,k}[t]=s_{b,k}[t]\exp(j2\pi f_c t),
	\end{equation}where $s_{b,k}[t]$ is the baseband waveform, $\displaystyle f_c=\frac{c}{\lambda}$ is the carrier frequency, and $\lambda$ is the wavelength. In this work, we only consider LoS propagation, as NLoS paths are strongly attenuated at high carrier frequencies.
	
	The received signal from the $b$-th anchor at the receiver's $u$-th antenna element during the $k$-th transmission slot is thus given as:
	\begin{equation}
		\label{eqn:rx_sig}
		y_{bu,k}[t]=\beta_{bu,k}s_{b,k}[\tilde{t}_{bu,k}]\exp\left(j2\pi {f}_{d,bu,k}\tilde{t}_{bu,k}\right)+w_{bu,k}[t],
	\end{equation}where $w_{bu,k}[t]\sim \mathcal{C}\mathcal{N}\left(0,N_0\right)$ is the thermal noise, $\beta_{bu,k}$ is the channel gain, $f_{d,bu,k}=f_c(1-\nu_{bu,k})+\epsilon_b$, is the observed Doppler frequency, and $\tilde{t}_{bu,k}=t-\tau_{bu,k}$, is the effective time duration between $b$-th anchor and the $u$-th receive antenna during the $k$-th transmission slot, respectively. Here, $\displaystyle\tau_{bu,k}=\frac{\|\mathbf{p}_{bu,k}\|}{c}+\delta_b$ is the observed propagation delay, $\nu_{bu,k}$ is the observed Doppler shift, $\delta_b$ is the unknown time offset between the receiver and the $b$-th anchor, and $\epsilon_b$ is the unknown frequency offset between the receiver and $b$-th anchor.
	
	For the channel gain, we use the Friis-space path loss model given as:
	\begin{equation}
		\beta_{bu,k}=\frac{\lambda}{4\pi d_{bu,k}}.
	\end{equation}
	
	\subsection{Doppler frequency}
	
	\begin{figure}[!htbp]
		\centering\includegraphics[clip,trim=3.5cm 6.8cm 1cm 2.8cm, width=1.45\linewidth, keepaspectratio]{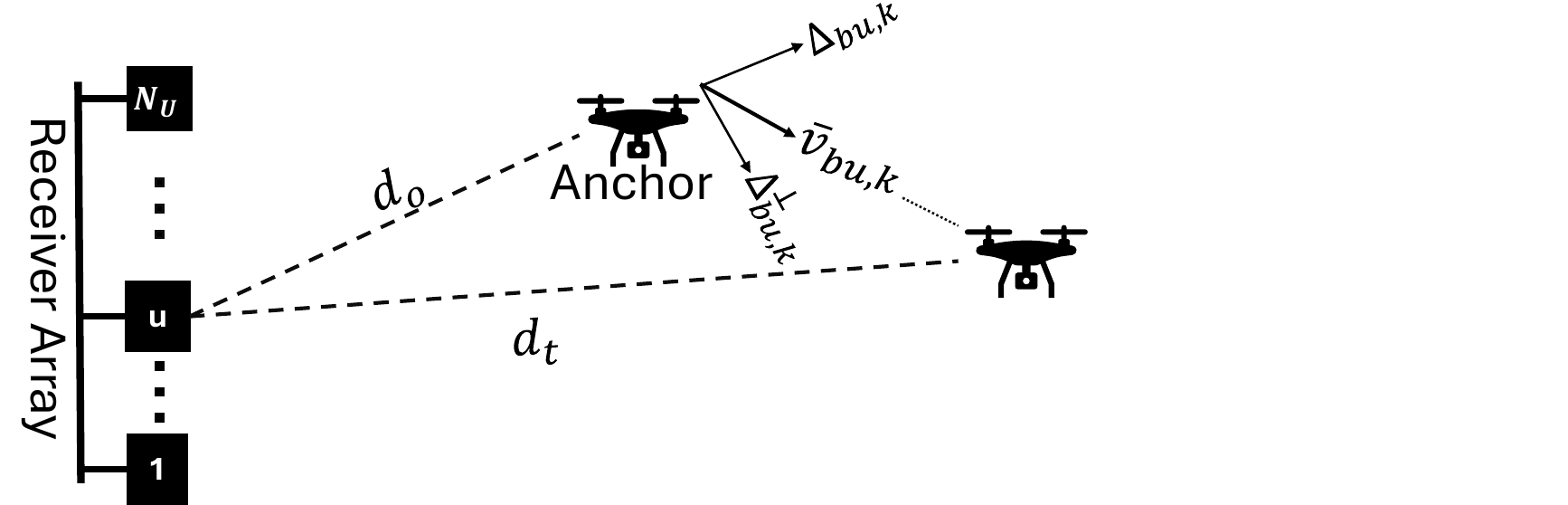}
		\caption{Illustration of the relative anchor movement with respect to the receiver array during the $k$-th transmission slot.} \label{fig:dopp_elaa}
	\end{figure}
	
	Estimating Doppler frequency is a key step for velocity sensing. Different wavefronts, planar in FF and spherical in NF, enable different estimation capabilities of velocity components. For instance, in FF, Doppler frequency is only affected by radial velocity component, while NF allows for full velocity estimation \cite{nf_vel_sen_pred_bf,perf_bound_vel_est_elaa}. The concept of velocity sensing in the NF is still in a nascent stage. For completeness, we derive the spherical-wave Doppler model that governs element dependent frequency shifts. 
    
    Fundamentally, observed frequency is defined as the instantaneous rate of change in phase of a received waveform. 
	We analyze the phase of the measured electric field component at any arbitrary receiver antenna element, which can be written as follows \cite{loc_sig_nf_aa}:
	\begin{equation}
		\label{eq:e_field_nf}
		E(f,t,d)=A\frac{\cos\left(2\pi f\left(t-\frac{d_t}{c}\right)\right)}{d_t},
	\end{equation}where $d_t$ is the propagation distance at time $t$, $f$ is the transmission frequency, and the constant $A$ takes care of phase independent scaling terms.
	
	\begin{remark}
		As highlighted in \cite{loc_sig_nf_aa}, the expressions for amplitude decay and phase are much more complex than they seem. However, for our targetted NF operation, {\em i.e.} in the Fresnel region, specified by $\displaystyle \left(0.62\sqrt{\frac{D^3}{\lambda}}\right)<d<\displaystyle \left(\frac{2 D^2}{\lambda}\right)$, the expression in ~\eqnref{eq:e_field_nf} holds. Here, $D$ denotes the largest dimension of the antenna array.
	\end{remark}
	
	\begin{remark}
		Let $T_{dp}$ denote the time duration of the transmission slot where pulses used for Doppler sensing are transmitted. Practically, these are very short-duration pulses in the order of nano-seconds, and given the distances, velocities, and the propagation regime we are operating in, for all practical considerations we can assume the radial direction $\Delta_{bu,k}$ remains constant during the pulse transmission slot. This yields a constant time-invariant Doppler shift. Therefore, we can write the propagation distance at time $t$ as:
	\end{remark}
	\begin{equation*}        d_t=d_o+\left(\mathbf{\bar{v}}_{bu,k}\cdot\Delta_{bu,k} \right)t, \;\;\; 0<t\leq T_{dp}.
	\end{equation*}The instantaneous observed phase at time $t$ is given as:
	\begin{equation*}
		\phi(t)=2\pi f \left(t-\frac{d_t}{c}\right). 
	\end{equation*} The observed frequency at time $t$ can be obtained as:
	\begin{equation*}
		\begin{split}
			f_{ins}&=\phi'(t)=2\pi f\left(1-\frac{1}{c} \frac{{\rm d} d_t}{{\rm d}t}\right) \\
			&=2\pi f\left(1-\frac{\mathbf{\bar{v}}_{bu,k}\cdot\Delta_{bu,k}}{c}\right).
		\end{split}
	\end{equation*}The observed Doppler shift can be obtained as:
	\begin{equation}
		\nu_{bu,k}=\frac{\mathbf{\bar{v}}_{bu,k}\cdot\Delta_{bu,k}}{c}.
	\end{equation}A key takeaway is that in the NF each element observes a distinct angle of incidence and thus the target velocity projects differently along the different incidence directions, leading to element-wise varying Doppler shifts. Which is different from the FF where a single Doppler shift applies across the entire antenna array due to the fact that all antenna elements are practically at the same distance from the target and thus observe the same angle of incidence (or equivalently same phase shift across the array).  The diversity of projections in the NF provides higher rank velocity information, enabling full $3$D velocity recovery and tighter estimation bounds. 
	
	\subsection{Received Signal Properties}
	The following signal parameters will be used to derive the analytical results in the upcoming sections.
	
	\subsubsection{Spectral waveform parameters}
	Key signal properties will be derived in terms of the Fourier transform of the transmitted baseband signal, which is given as:
	
	\begin{equation*}
		S_{bu,k}[f]=\frac{1}{\sqrt{2\pi}} \int_{-\infty}^\infty s_{bu,k}[t] \exp(-j 2\pi ft) {\rm d}t.
	\end{equation*}
	
	\begin{enumerate}[label=(\alph*)]
		\item Received signal to noise ratio (SNR): The SNR quantifies the ratio of power of the signal across its occupied frequencies to the noise spectral density and is given as: 
		\begin{equation*}
			{\rm SNR}_{bu,k}=\frac{|\beta_{bu,k}|^2}{\sigma_w^2} \int_{-\infty}^\infty \left|S_{bu,k}[f]\right|^2 {\rm d}f.
		\end{equation*}
		\item Effective baseband bandwidth: This quantity captures the spread of occupied frequencies in the transmitted waveform, given as:
		\begin{equation*}
			\alpha_{1,bu,k}=\left(\frac{\int_{-\infty}^\infty f^2\left|S_{bu,k}[f]\right|^2 {\rm d}f}{\int_{-\infty}^\infty \left|S_{bu,k}[f]\right|^2 {\rm d}f}\right)^{\frac{1}{2}}.
		\end{equation*}
		\item Baseband carrier correlation (BCC): This quantity reflects how strongly the received spectrum is concentrated around its center frequency, given as:
			\begin{equation*}
            \begin{split}
				\alpha_{2,bu,k}&= \\
                &
				\frac{\int_{-\infty}^\infty f\left|S_{bu,k}[f]\right|^2 {\rm d}f}{\left(\int_{-\infty}^\infty f^2\left|S_{bu,k}[f]\right|^2 {\rm d}f\right)^{\frac{1}{2}}\left(\int_{-\infty}^\infty \left|S_{bu,k}[f]\right|^2 {\rm d}f\right)^{\frac{1}{2}}}.
                \end{split}
			\end{equation*}
	\end{enumerate}
	\subsubsection{Temporal waveform parameters}
	The following time domain metrics further determine how the transmitted waveform interacts with the NF channel and they are given as:
	\begin{enumerate}[label=(\alph*)]
		\item Effective pulse duration: This quantity captures the time spread (root mean square duration) of the energy in the baseband pulse and is given as:
		\begin{equation*}
			\sigma = \sqrt{\frac{\int_{-\infty}^\infty t^2 \left|s(t)\right|^2 {\rm d}t}{\int_{-\infty}^\infty \left|s(t)\right|^2 {\rm d}t}}.
		\end{equation*}
		\item Temporal-energy centroid: This quantity captures the energy-weighted time center of the pulse, given as:
		\begin{equation*}
			\gamma = \int_{-\infty}^\infty t \left|s(t)\right|^2 {\rm d}t.
		\end{equation*}For symmetric pulses, it evaluates to $0$.
		\item Correlation term: This quantity captures the correlation of a pulse with its first-order derivative. It is a boundary term that determines if the net pulse has a rising or decaying envelope. 
		\begin{equation*}
			\varepsilon = \int_{-\infty}^\infty s(t) s'(t) {\rm d}t=\frac{1}{2}\left[s^2(\infty)-s^2(-\infty)\right].
		\end{equation*}For finite-time signals, it evaluates to $0$.
	\end{enumerate}
	\section{Available Information in Received Signal}
	
	
	In this section, we derive the C-CRB for the estimation of $8$D position parameters in the presence of asynchronization and unknown channel gains. We first compute the FIM for the observable channel quantities (delays, Doppler, gains) and map it to the $8$D parameters via an appropriate Jacobian construction. Nuisance terms are then marginalized through the EFIM, quantifying the information loss due to asynchronization and unknown parameters (channel gains). Subsequently, we enforce the unit-norm constraint on the orientation vector to yield the C-CRB, providing tighter, physically consistent bounds. We identify the parameter tradeoffs and through an invertibility analysis, determine the minimal infrastructure required for $8$D localizability.
	\begin{figure*}[!htbp]
		\begin{equation}
			\label{eqn:likelihood}
			\mathcal{L}\left(\mathbf{y}[t] | \boldsymbol{\eta}\right) \propto \prod_{b=1}^{N_B}\prod_{k=1}^{N_K}\prod_{u=1}^{N_U}\exp\left\{-\frac{1}{N_o}\int_0^{T_{obs}}\left|y_{bu,k}[t]-\mu_{bu,k}[t]\right|^2{\rm d}t\right\}.
		\end{equation}
	\end{figure*}
	
	\subsection{Channel parameters}
	
	We begin by describing the geometric and nuisance channel parameters observed across the $N_U$ receive antennas during the $N_K$ transmission slots from all of the $N_B$ anchors.
	The delays from the $b$-th anchor during the $k$-th transmission slot across all the receiver antennas are vectorized as:
	\begin{equation*}
		\boldsymbol{\tau}_{b,k}=\left[\tau_{b1,k},\tau_{b2,k},\dots,\tau_{b N_U,k}\right]^{\rm T}.
	\end{equation*}
	Similarly, the observed Doppler from the $b$-th anchor during the $k$-th transmission slot across all the receiver antennas is vectorized as:
	\begin{equation*}
		\boldsymbol{f}_{d,b,k}=\left[f_{d,b1,k},f_{d,b2,k},\dots,f_{d,b N_U,k}\right]^{\rm T}.
	\end{equation*}
	Similarly, the channel gain observed from the $b$-th anchor during the $k$-th transmission slot across all the receiver antennas is vectorized as:
	\begin{equation*}
		\boldsymbol{\beta}_{b,k}=\left[\beta_{b1,k},\beta_{b2,k},\dots,\beta_{b N_U,k}\right]^{\rm T}.
	\end{equation*}We further vectorize them across the time slots for the $b$-th anchor as follows:
	\begin{equation*}
		\begin{split}
			\boldsymbol{\tau}_{b}&=\left[\boldsymbol{\tau}_{b,1}^{\rm T},\boldsymbol{\tau}_{b,2}^{\rm T},\dots,\boldsymbol{\tau}_{b,k}^{\rm T}\right]^{\rm T}, \\
			\boldsymbol{f}_{d,b}&=\left[\boldsymbol{f}_{d,b,1}^{\rm T},\boldsymbol{f}_{d,b,2}^{\rm T},\dots,\boldsymbol{f}_{d,b,k}^{\rm T}\right]^{\rm T}, \\
			\boldsymbol{\beta}_{b}&=\left[\boldsymbol{\beta}_{b,1}^{\rm T},\boldsymbol{\beta}_{b,2}^{\rm T},\dots,\boldsymbol{\beta}_{b,k}^{\rm T}\right]^{\rm T}.
		\end{split}
	\end{equation*} Next, the observable parameters from the $b$-th anchor including the nuisance offsets are vectorized as:
	\begin{equation*}
		\boldsymbol{\eta}_b=\left[\boldsymbol{\tau}_b^{\rm T},\boldsymbol{f}_{d,b}^{\rm T},\boldsymbol{\beta}_b^{\rm T},\delta_b,\epsilon_b\right]^{\rm T}.
	\end{equation*} Finally, we aggregate all the observable geometric and nuisance parameters as:
	\begin{equation*}
		\boldsymbol{\eta}=\left[\boldsymbol{\eta}_1^{\rm T},\boldsymbol{\eta}_2^{\rm T},\dots,\boldsymbol{\eta}_{N_B}^{\rm T}\right]^{\rm T}.
	\end{equation*}Given the parameter set $\boldsymbol{\eta}$, the likelihood of observing the received signal is given in ~\eqnref{eqn:likelihood}.
	
	\subsection{Mathematical preliminaries}
	The CRLB is a fundamental lower bound on the variance of any unbiased estimator. Mathematically, it is defined as a function of the inverse of the FIM. The FIM determines the information content about parameters of interest ($\boldsymbol{\eta}$) in realizations of a random variable with a given distribution $\displaystyle\left(\mathcal{L}\left(\mathbf{y}[t] | \boldsymbol{\eta}\right)\right)$. The invertibility of the FIM determines whether it is possible to estimate a parameter from a set of available observations $\mathbf{y}[t]$.
	\begin{definition}
		The FIM obtained from the observations in ~\eqnref{eqn:rx_sig}, with the likelihood function given in ~\eqnref{eqn:likelihood}, is defined as follows:
		\begin{equation}
			\mathbf{\rm J}_{\mathbf{y};\boldsymbol{\eta}}=-\mathbb{E}\left[\frac{\partial^2 {\rm ln}\; \mathcal{L}\left(\mathbf{y} ; \boldsymbol{\eta}\right)}{\partial\boldsymbol{\eta}\partial\boldsymbol{\eta}^T}\right].
		\end{equation}
		The CRLB for any unbiased estimator of $\boldsymbol{\eta}$ is thus given as:
		\begin{equation}
			{\rm CRLB}(\boldsymbol{\eta})={\rm Tr}\left(\mathbf{\rm J}_{\mathbf{y};\boldsymbol{\eta}}^{-1}\right).
		\end{equation}
		It is important to note here that the size of the FIM grows quadratically with the number of parameters; hence, the complexity of inversion also grows quadratically. 
	\end{definition}
	\begin{definition}
		When it is desired to estimate only a subset of parameters $\boldsymbol{\eta}_1$ from $\boldsymbol{\eta}=\left[\boldsymbol{\eta}_1,\boldsymbol{\eta}_2\right]$ where $\boldsymbol{\eta}_2$ denotes the nuisance parameters, the resultant FIM takes the following structure:
		\begin{equation*}
			\mathbf{\rm J}_{\mathbf{y};\boldsymbol{\eta}} = \left[\begin{array}{cc}
				\mathbf{\rm J}_{\mathbf{y};\boldsymbol{\eta}_1}&  \mathbf{\rm J}_{\mathbf{y};\boldsymbol{\eta}_1,\boldsymbol{\eta}_2}\\
				\mathbf{\rm J}_{\mathbf{y};\boldsymbol{\eta}_1,\boldsymbol{\eta}_2}^{\rm T} & \mathbf{\rm J}_{\mathbf{y};\boldsymbol{\eta}_2}
			\end{array}\right].
		\end{equation*}The equivalent FIM (EFIM) for the parameters $\boldsymbol{\eta}_1$ is defined as:
		\begin{equation*}
			\mathbf{\rm J}_{\mathbf{y};\boldsymbol{\eta}_1}^{e}=\mathbf{\rm J}_{\mathbf{y};\boldsymbol{\eta}_1}-\mathbf{\rm J}_{\mathbf{y};\boldsymbol{\eta}_1,\boldsymbol{\eta}_2}\mathbf{\rm J}_{\mathbf{y};\boldsymbol{\eta}_2}^{-1}\mathbf{\rm J}_{\mathbf{y};\boldsymbol{\eta}_1,\boldsymbol{\eta}_2}^{\rm T}.
		\end{equation*} Here, the latter term denotes the information loss due to unknown nuisance parameters. The EFIM captures all the information present in the FIM about the parameter of interest and follows the following information equality:
		\begin{equation*}
			\left(\mathbf{\rm J}_{\mathbf{y};\boldsymbol{\eta}_1}^{e}\right)^{-1}=\left[\mathbf{\rm J}_{\mathbf{y};\boldsymbol{\eta}}^{-1}\right]_{\left[1:n,1:n\right]}.
		\end{equation*}
	\end{definition}
	
	\subsection{Available channel information}
	
	In this section, we derive the FIM, which defines the available information about the channel parameters from the received signals. We begin by defining:
	\begin{equation}
		\mathbf{F}_{\mathbf{y}}\left(\mathbf{y}|\boldsymbol{\eta};\theta_1,\theta_2\right)=-\mathbb{E}\left[\frac{\partial^2 {\rm ln}\; \mathcal{L}\left(\mathbf{y} ; \boldsymbol{\eta}\right)}{\partial\theta_1\partial\theta_2}\right],
	\end{equation} where $\theta_1$ and $\theta_2$ are any two channel parameters. Assuming the signals received from different anchors are independent across time slots and antenna elements, we obtain the following block diagonal matrix:
	\begin{equation}
		\begin{split}               &\mathbf{F}_{\mathbf{y}}\left(\mathbf{y}|\boldsymbol{\eta};\boldsymbol{\eta},\boldsymbol{\eta}\right) = {\rm diag}\left\{  \mathbf{F}_{\mathbf{y}}\left(\mathbf{y}|\boldsymbol{\eta};\boldsymbol{\eta}_1,\boldsymbol{\eta}_1\right) , \right.\\
			&\left.\mathbf{F}_{\mathbf{y}}\left(\mathbf{y}|\boldsymbol{\eta};\boldsymbol{\eta}_2,\boldsymbol{\eta}_2\right)  , \dots, \mathbf{F}_{\mathbf{y}}\left(\mathbf{y}|\boldsymbol{\eta};\boldsymbol{\eta}_{N_B},\boldsymbol{\eta}_{N_B}\right)   \right\}.
		\end{split}
	\end{equation}
	Using the likelihood function given in ~\eqnref{eqn:likelihood}, we obtain the following simplified expression:
	\begin{equation}
		\begin{split}
			&\mathbf{F}_{\mathbf{y}}\left(\mathbf{y}|\boldsymbol{\eta}_b;\theta_1,\theta_2\right) = \\
			&\frac{2}{N_o}\sum_{k=1}^{N_K} \sum_{u=1}^{N_U} {\rm Re}\left\{\int_{T_k}^{T_k+T_{ob}}\frac{\partial \mu_{bu,k}[t]}{\partial \theta_1}\frac{\partial\mu_{bu,k}^{\rm H} [t]}{\partial \theta_2}\right\}.
		\end{split}
	\end{equation}The individual FIM entries can be obtained as:
	
	\begin{equation}
		\label{eqn:fim_entry}
		\begin{split}
			&\mathbf{F}_{\mathbf{y}}\left(\mathbf{y}|\boldsymbol{\eta}_b;\tau_{bu,k},\tau_{bu,k}\right)= \\
			&\;\;\;\;\;\;
			\frac{8\pi^2\left|\beta_{bu,k}\right|^2 E_S}{N_o}\left[{f}_{d,bu,k}^2 +\alpha_{1,bu,k}^2+{f}_{d,bu,k}\alpha_{2,bu,k}\right], \\
			&\mathbf{F}_{\mathbf{y}}\left(\mathbf{y}|\boldsymbol{\eta}_b;f_{d,bu,k},f_{d,bu,k}\right)=\frac{8\pi^2  \left|\beta_{bu,k}\right|^2}{N_o}\sigma^2, \\
			&\mathbf{F}_{\mathbf{y}}\left(\mathbf{y}|\boldsymbol{\eta}_b;\beta_{bu,k},\beta_{bu,k}\right)=\frac{1}{4\pi^2\left|\beta_{bu,k}\right|^2}{\rm SNR}_{bu,k} ,\\
			&\mathbf{F}_{\mathbf{y}}\left(\mathbf{y}|\boldsymbol{\eta}_b;\beta_{bu,k},\tau_{bu,k}\right)=\frac{-2}{N_o}\left|\beta_{bu,k}\right| \varepsilon, \\
			&\mathbf{F}_{\mathbf{y}}\left(\mathbf{y}|\boldsymbol{\eta}_b;\tau_{bu,k},f_{d,bu,k}\right)=\frac{8\pi^2  \left|\beta_{bu,k}\right|^2}{N_o}f_c f_{d,bu,k}\gamma, \\
			&\mathbf{F}_{\mathbf{y}}\left(\mathbf{y}|\boldsymbol{\eta}_b;\delta_{b},\delta_{b}\right)=\sum_{u=1}^{N_u} \sum_{k=1}^{N_k} \mathbf{F}_{\mathbf{y}}\left(\mathbf{y}|\boldsymbol{\eta};\tau_{bu,k},\tau_{bu,k}\right),\\ 
			&\mathbf{F}_{\mathbf{y}}\left(\mathbf{y}|\boldsymbol{\eta}_b;\epsilon_{b},\epsilon_{b}\right)= \sum_{u=1}^{N_u} \sum_{k=1}^{N_k} \mathbf{F}_{\mathbf{y}}\left(\mathbf{y}|\boldsymbol{\eta};f_{d,bu,k},f_{d,bu,k}\right),\\
			&\mathbf{F}_{\mathbf{y}}\left(\mathbf{y}|\boldsymbol{\eta}_b;\delta_{b},\epsilon_{b}\right)= \sum_{u=1}^{N_u} \sum_{k=1}^{N_k} \mathbf{F}_{\mathbf{y}}\left(\mathbf{y}|\boldsymbol{\eta};\tau_{bu,k},f_{d,bu,k}\right),\\
			&\mathbf{F}_{\mathbf{y}}\left(\mathbf{y}|\boldsymbol{\eta}_b;\tau_{bu,k},\delta_{b}\right)=\mathbf{F}_{\mathbf{y}}\left(\mathbf{y}|\boldsymbol{\eta};\tau_{bu,k},\tau_{bu,k}\right), \\
			&\mathbf{F}_{\mathbf{y}}\left(\mathbf{y}|\boldsymbol{\eta}_b;f_{d,bu,k},\delta_{b}\right)=\mathbf{F}_{\mathbf{y}}\left(\mathbf{y}|\boldsymbol{\eta};\tau_{bu,k},f_{d,bu,k}\right), \\
			&\mathbf{F}_{\mathbf{y}}\left(\mathbf{y}|\boldsymbol{\eta}_b;\beta_{bu,k},\delta_{b}\right)=\mathbf{F}_{\mathbf{y}}\left(\mathbf{y}|\boldsymbol{\eta};\beta_{bu,k},\tau_{bu,k}\right),\\    
			&\mathbf{F}_{\mathbf{y}}\left(\mathbf{y}|\boldsymbol{\eta}_b;\tau_{bu,k},\epsilon_{b}\right)=\mathbf{F}_{\mathbf{y}}\left(\mathbf{y}|\boldsymbol{\eta};\tau_{bu,k},f_{d,bu,k}\right), \\
			&\mathbf{F}_{\mathbf{y}}\left(\mathbf{y}|\boldsymbol{\eta}_b;f_{d,bu,k},\epsilon_{b}\right)=\mathbf{F}_{\mathbf{y}}\left(\mathbf{y}|\boldsymbol{\eta};f_{d,bu,k}, f_{d,bu,k}\right), \\
			&\mathbf{F}_{\mathbf{y}}\left(\mathbf{y}|\boldsymbol{\eta}_b;\beta_{bu,k},f_{d,bu,k}\right)=\mathbf{F}_{\mathbf{y}}\left(\mathbf{y}|\boldsymbol{\eta};\beta_{bu,k}, \epsilon_{b}\right)=0.
		\end{split}
	\end{equation}
	\subsection{Conversion to useful $8$D position parameters}
	
	In this section, we describe the $8$D position parameters of interest and derive mathematical representations of their information content in the FIM.
	
	\begin{definition}
		The FIM under a bijective function mapping one parameter space $\boldsymbol{\eta}=\left[\eta_1,\eta_2,\dots,\eta_n\right]$ to another $\boldsymbol{\kappa}=\left[\kappa_1,\kappa_2,\dots,\kappa_k\right]$ gets transformed as follows \cite{handbook_of_loc}:
		\begin{equation*}
			\mathbf{\rm J}_{\mathbf{y};\boldsymbol{\kappa}}=\mathbf{J}_{\boldsymbol{\eta}\rightarrow \boldsymbol{\kappa}}\mathbf{\rm J}_{\mathbf{y};\boldsymbol{\eta}}\mathbf{J}_{\boldsymbol{\eta}\rightarrow \boldsymbol{\kappa}}^{\rm T},
		\end{equation*} where $\mathbf{J}_{\boldsymbol{\eta}\rightarrow \boldsymbol{\kappa}}$ is a transformation matrix capturing the non-linear relationship between the parameters in different spaces, given as:
		\begin{equation*}
			\mathbf{J}_{\boldsymbol{\eta}\rightarrow \boldsymbol{\kappa}}=\left[\begin{array}{cccc}
				\frac{\partial \eta_1}{\partial \kappa_1} & \frac{\partial \eta_2}{\partial \kappa_1} & \dots & \frac{\partial \eta_n}{\partial \kappa_1}  \\
				
				\vdots & \vdots & \ddots & \vdots \\
				
				\frac{\partial \eta_1}{\partial \kappa_k} & \frac{\partial \eta_2}{\partial \kappa_k} & \dots & \frac{\partial \eta_n}{\partial \kappa_k}				
			\end{array}\right].
		\end{equation*}
	\end{definition}
	
	The location parameters that are of interest are:
	\begin{enumerate}
		\item The $3$D position of the receiver at the beginning of the zero-th time slot, {\em i.e.}, $\mathbf{p}_{U,0}$.
		\item The $3$D velocity of receiver $\mathbf{\bar{v}}_U$ which is assumed constant across the $N_K$ time slots.
		\item The receiver's $3$D orientation vector $\bar{\mathbf{s}}$ which is assumed constant across the $N_K$ time slots.
	\end{enumerate}
	Lastly, the nuisance parameters that contribute to the information loss are the unknown time offsets $\left\{\delta_b\right\}_{b=1}^{N_B}$, frequency offsets $\left\{\epsilon_b\right\}_{b=1}^{N_B}$, and channel gains $\left\{\beta_{bu,k}\right\}_{\forall b,u,k}$. Thus, all the unknown parameters can be vectorized as:
	\begin{equation*}
		\boldsymbol{\kappa}=\left[\boldsymbol{\kappa}_1^{\rm T},\boldsymbol{\kappa}_2^{\rm T}\right]^{\rm T},
	\end{equation*}where $\boldsymbol{\kappa}_1=\left[\mathbf{p}_{U,0}^{\rm T},\mathbf{\bar{v}}_U^{\rm T},\bar{\mathbf{s}}^{\rm T}\right]^{\rm T}$ denotes the position parameters of interest and $\boldsymbol{\kappa}_2=\left[\left\{\delta_b\right\}_{b=1}^{N_B},\left\{\epsilon_b\right\}_{b=1}^{N_B},\left\{\beta_{bu,k}\right\}_{\forall b,u,k}\right]^{\rm T}$ denotes the nuisance parameters. 
	The entries for the transformation matrix $\mathbf{J}_{\boldsymbol{\eta}\rightarrow \boldsymbol{\kappa}}$ are derived in Appendix~\ref{appn:A}. 
	\subsection{Constrained CRB (C-CRB)}
	To obtain tighter bounds, we enforce the unit-norm constraint on the orientation vector $\bar{\mathbf{s}}$, {\em i.e,} $\|\bar{\mathbf{s}}\|=1$. The constraint vector can be written as: 
	\begin{equation}
		\begin{split}
			\mathbf{h}(\boldsymbol{\kappa}_1) &= \left[
			\|\bar{\mathbf{s}}\|^2 - 1\
			\right]
			= 0.
		\end{split}
	\end{equation}
	The C-CRB gives the lower bound on the variance of any unbiased estimator subject to parametric constraints and can be obtained as \cite{stoica1998cramer}:
	\begin{equation}
		\label{eqn:ccrb}
		{\rm CCRB}(\boldsymbol{\kappa}_1)=\mathbf{U}\left(\mathbf{U}^{\rm T} \mathbf{J}_{\boldsymbol{\kappa}_1}\mathbf{U}\right)^{-1}\mathbf{U}^{\rm T},
	\end{equation}
	where $\mathbf{U}$ is a matrix whose columns form an orthonormal basis for the null space of $\displaystyle \mathbf{H}=\left[\frac{\partial \mathbf{h}^{\rm T}}{\partial\boldsymbol{\kappa}_1}\right]$. 
	This formulation effectively projects the FIM onto the tangent space of the constraint manifold, ensuring that only perturbations consistent with the imposed constraints contribute to the bound.
	
	
	\section{$8$D Localization Algorithm} 
In this section, we present a two-stage localization approach for estimating $8$D position parameters and offsets. We divide our approach into channel parameter estimation and ML based $8$D position inference.
\subsection{Channel parameter estimation}
The channel parameters to be estimated for our localization problem include the delays (ToA) and Doppler frequencies observed at the individual ELAA elements. Since channel parameter estimation has been a rigorously studied topic in literature, for the sake of conciseness we do not discuss that in this work. There are different algorithms such as RiMAX \cite{thoma2004rimax} and SAGE \cite{fleury1999channel} for estimating Doppler frequency. Approaches for ToA estimation include fast Fourier transform (FFT) based methods \cite{SANTAMARIA2000819}. ToA estimation for UWB signals has been discussed in \cite{5508975}. Matched filters are another very popular approach for ToA estimation \cite{wu2008match}. Approach for joint detection of Doppler frequency and delay estimation is presented in \cite{wang2024joint,7294690}. CRLB or near-CRLB convergence of the channel parameter estimate is demonstrated in \cite{fleury1999channel,7294690}.

We assume an arbitrary estimator is applied to estimate the useful channel parameters $\boldsymbol{\hat{\eta}}=\left[\boldsymbol{\hat{\tau}_b}^{\rm T},\mathbf{\hat{f}_{d,b}}^{\rm T}\right]$. For delay measurements, we assume a Gaussian distribution:
\begin{equation}
	p(\hat{\tau}_{bu,k}|\tau_{bu,k})=\frac{1}{\sqrt{2\pi\sigma_\tau^2}}\exp\left(-\frac{1}{2\sigma_\tau^2}\left(\hat{\tau}_{bu,k}-\tau_{bu,k}\right)^2\right),
\end{equation}
where $\sigma_\tau$ denotes the error of the delay estimator used.

For Doppler frequency measurements, we also assume a Gaussian distribution:
	\begin{equation}
		\begin{split}p(\hat{f}_{d,bu,k}|f_{d,bu,k})&=
        \\& \frac{1}{\sqrt{2\pi\sigma_d^2}}\exp\left(-\frac{1}{2\sigma_d^2}\left(\hat{f}_{d,bu,k}-f_{d,bu,k}\right)^2\right),
        \end{split}
	\end{equation}
where $\sigma_d$ denotes the  error of the Doppler frequency estimator used.

\subsection{ML estimation}
Our goal is to estimate the $8$D position parameters and unknown offsets vectorized as $\boldsymbol{\chi}=\left[\boldsymbol{\kappa}_1^{\rm T}, \left\{\delta_b,\epsilon_b\right\}_{\forall b}\right]^T$. We begin by formulating the negative joint log likelihood cost function as:
\begin{equation}
	\begin{split}
		\mathcal{L}\left(\bar{\mathbf{s}},\boldsymbol{\chi}\right)&=\sum_{b=1}^{N_B}\sum_{u=1}^{N_U}\sum_{k=1}^{N_K}\frac{1}{2\sigma_\tau^2}\left(\hat{\tau}_{bu,k}-\tau_{bu,k}\left(\bar{\mathbf{s}},\boldsymbol{\chi}\right)\right)^2 + \\
		&\frac{1}{2\sigma_d^2}\left(\hat{f}_{d,bu,k}-f_{d,bu,k}\left(\bar{\mathbf{s}},\boldsymbol{\chi}\right)\right)^2.
	\end{split}
\end{equation}
Using the above-defined cost function, the ML estimation can be formulated as the following optimization problem:

\begin{equation}
	\label{eq:opt_prob}
	\begin{aligned}
		&\min_{\bar{\mathbf{s}},\boldsymbol{\chi}} \quad \mathcal{L}\left(\bar{\mathbf{s}},\boldsymbol{\chi}\right) \\
		&{\rm s.t} \quad \bar{\mathbf{s}}^{\rm T}\bar{\mathbf{s}}=1.
	\end{aligned}
\end{equation}
Here, we note that the above problem is non-convex and non-linear with respect to the optimization variables. We propose to solve this iteratively by breaking it into two simpler subproblems.
\subsubsection{Constrained optimization for orientation estimation}
We apply Riemannian gradient descent (RGD) to optimize $\bar{\mathbf{s}}$ subject to the constraint in ~\eqnref{eq:opt_prob}
as: \cite{boumal2023introduction}
\begin{equation}
	\label{eq:rgd}
	\hat{\bar{\mathbf{s}}}^{(k+1)}=\mathcal{R}_{\bar{\mathbf{s}}^{(k)}}\left(-\varepsilon_k\mathcal{P}_{\bar{\mathbf{s}}^{(k)}}\left(\left.\frac{\partial\mathcal{L}}{\partial\bar{\mathbf{s}}}\right)\right|_{\hat{\bar{\mathbf{s}}}^{(k)},\boldsymbol{\chi}^{(k)}}\right),
\end{equation}
where $\varepsilon_k$ is the step size, $\mathcal{P}_x(u)$ denotes the orthogonal projection of $u$ on the tangent space of $x$, and $\mathcal{R}_x(u)$ is the retraction operator from the tangent space back onto the constraint manifold. These operators are defined as\cite{boumal2023introduction}:
\begin{align}
	\mathcal{P}_x(u)&= u-\langle x,u\rangle x, \\
	\mathcal{R}_x(u)&=\frac{x+u}{\sqrt{1+\|u\|^2}}.
\end{align}
\subsubsection{Block-wise optimization for position and velocity estimation}
We employ the line search method to optimize $\boldsymbol{\chi}$ which lies in the unconstrained Euclidean space \cite{nocedal1999numerical}. To improve stability, we adopt a block–coordinate descent inspired scheme that cyclically updates position, velocity, and offset variables while holding the others fixed. Because the parameters and gradients operate on very different numerical scales (e.g., Doppler related terms are of the order of GHz, whereas clock offsets are of the order of microseconds), each block uses its own step size. This reduces ill-conditioning and cross-block gradient interference, yielding faster, more reliable convergence than a single joint update. The complete methodology for ML estimation is given in Algorithm~\ref{alg:1}.

\begin{algorithm}
	\caption{ML refinement}
	\label{alg:1}
	\begin{algorithmic}[1]
		\State \textbf{Input:} Delay and Doppler measurements $\left\{\hat{\tau}_{bu,k},\hat{f}_{d,bu,k}\right\}_{\forall b,u,k}$.
		\State Obtain initial estimates $\hat{\boldsymbol{\chi}}^{(0)}$ and $\hat{\bar{\mathbf{s}}}^{(0)}$ using Algorithm-2.
		\While{Stopping criteria is not met}
		\Statex $\boldsymbol{\circ}$ {\bf Fix:} $\hat{\bar{\mathbf{s}}}^{(k)}$, $\hat{\bar{\mathbf{v}}}_U^{(k)}$, $\left\{\hat{\delta}_b^{(k)}, \hat{\epsilon}_b^{(k)}\right\}_{b=1}^{N_B}$; {\bf Update}: $\hat{\mathbf{p}}_{U,0}^{(k)}$ using line search.
		\Statex $\boldsymbol{\circ}$ {\bf Fix:} $\hat{\bar{\mathbf{s}}}^{(k)}$, $\hat{\mathbf{p}}_{U,0}^{(k)}$, $\left\{\hat{\delta}_b^{(k)}, \hat{\epsilon}_b^{(k)}\right\}_{b=1}^{N_B}$; {\bf Update}: $\hat{\bar{\mathbf{v}}}_U^{(k)}$ using line search.
		\Statex $\boldsymbol{\circ}$ {\bf Fix:} $\hat{\bar{\mathbf{s}}}^{(k)}$, $\hat{\mathbf{p}}_{U,0}^{(k)}$, $\hat{\bar{\mathbf{v}}}_U^{(k)}$; {\bf Update}: $\left\{\hat{\delta}_b^{(k)}, \hat{\epsilon}_b^{(k)}\right\}_{b=1}^{N_B}$ using line search.
		\Statex $\boldsymbol{\circ}$ {\bf Fix:} $\hat{\mathbf{p}}_{U,0}^{(k)}$, $\hat{\bar{\mathbf{v}}}_U^{(k)}$, $\left\{\hat{\delta}_b^{(k)}, \hat{\epsilon}_b^{(k)}\right\}_{b=1}^{N_B}$; {\bf Update}: $\hat{\bar{\mathbf{s}}}^{(k)}$ using ~\eqnref{eq:rgd}.
		\EndWhile
		\State \textbf{Output:} $\hat{\boldsymbol{\chi}}$ and $\hat{\bar{\mathbf{s}}}$.
	\end{algorithmic}
\end{algorithm}


\subsection{Geometric initializer}
The ML refinement algorithm needs a good initialization to ensure it reaches a global optimum. Otherwise, due to the problem's non-convexity, poor initializations may lead to inefficient local optima. In this section, we propose a generic approach to design a simple initializer that leverages the array's linear geometry and system's rectilinear kinematics to come up with a crude estimate that will serve as the initial guess for the ML estimation procedure.

There are a variety of approaches in the literature for delay-based localization \cite{handbook_of_loc}. We can choose any convenient approach and begin by estimating the positions of an arbitrary number of antenna elements across different time slots. We can denote the set of indices of antenna elements by $\mathcal{I}\subset\{1,\dots,N_U\}$. A larger index set will produce a more precise initializer. We let $\hat{\mathbf{p}}_{u,k}$, which is obtained from the localization technique, represent the position of the $u\in\mathcal{I}$ antenna element during the $k$-th time slot. 

We utilize the ELAA’s linear geometry and system's rectilinear kinematics to obtain an initial estimate of the positioning parameters, which is then provided to the ML optimization module for further refinement. We note that Doppler frequency based least-squares (LS) estimation performs poorly with large frequency offsets. Therefore, we use it for velocity estimation only in the single snapshot case. The detailed methodology is presented in Algorithm~\ref{alg:2}. Note that, $\mathbf{stack}(.)$ and $\mathbf{vectorize}(.)$ operation gives the following vector and matrix, respectively:
\begin{equation}
\begin{split}
    \bar{\mathbf{f}_{\rm d}}\left((b-1)\times N_U+u\right)&= 1-\frac{\hat{f}_{d,bu}}{f_c},\\
    \mathbf{E}\left((b-1)\times N_U+u,1:3\right)&=\frac{\hat{\mathbf{p}}_{u,k}-\mathbf{p}^o_{b,k}}{\|\hat{\mathbf{p}}_{u,k}-\mathbf{p}^o_{b,k}\|}.
    \end{split}
\end{equation}

\begin{algorithm}
	\caption{Geometric initializer}
	\label{alg:2}
	\begin{algorithmic}[1]
		\State \textbf{Input:} $\{\hat{\mathbf{p}}_{u,k}\}_{\forall k, u\in \mathcal{I}}, \mathcal{I}$
		\State \textbf{Init:} $\hat{\bar{\mathbf{s}}}=\mathbf{0}_3$, $\hat{\mathbf{p}}_{0}=\mathbf{0}_3$, and $\hat{\mathbf{v}}_u=\mathbf{0}_3$.
		\State \textbf{Calculate} $\hat{\bar{\mathbf{s}}}:$
		\Statex \;\;\;\; \textbf{for} $k=1:N_K$
		\Statex \;\;\;\;\;\;\;\;\textbf{for} $i$ in $\mathcal{I}$
		\Statex \;\;\;\;\;\;\;\;\;\;\;\; $\displaystyle \hat{\bar{\mathbf{s}}}=\hat{\bar{\mathbf{s}}}+\frac{\hat{\mathbf{p}}_{i,k}-\hat{\mathbf{p}}_{i_{\rm prev},k}}{N_K \left|\mathcal{I}\right|}$ 
		\Statex \;\;\;\;\;\;\;\;\textbf{endfor}
		\Statex \;\;\;\;\textbf{endfor}
		\State \textbf{Calculate} $\hat{\mathbf{v}}_{u}:$
		\Statex \;\;\;\;\textbf{if} $N_k==1$
		\Statex\;\;\;\;\;\;\;\; $\bar{\mathbf{f}}_d= \mathbf{vectorize}\left(1-\frac{\hat{f}_{d,bu}}{f_c}\right)_{\forall b,u \in \mathcal{I}}$
		\Statex\;\;\;\;\;\;\;\; $\mathbf{E}= \mathbf{stack}\left[\frac{\hat{\mathbf{p}}_{u,k}-\mathbf{p}^o_{b,k}}{\|\hat{\mathbf{p}}_{u,k}-\mathbf{p}^o_{b,k}\|}\right]_{\forall b,u \in \mathcal{I}}$
		\Statex\;\;\;\;\;\;\;\; $\hat{\mathbf{v}}_{u}=\left(\mathbf{E}^{\rm T}\mathbf{E}\right)^{-1}\mathbf{E}^{\rm T}\bar{\mathbf{f}}_d$
		\Statex \;\;\;\;\textbf{else}
		\Statex \;\;\;\;\;\;\;\; \textbf{for} $k=2:N_K$
		\Statex \;\;\;\;\;\;\;\;\;\;\;\;\textbf{for} $i$ in $\mathcal{I}$
		\Statex \;\;\;\;\;\;\;\;\;\;\;\;\;\;\;\; $\hat{\mathbf{v}}_{u}=\hat{\mathbf{v}}_{u}+\frac{\hat{\mathbf{p}}_{i,k}-\hat{\mathbf{p}}_{i,k-1}}{(N_K-1) \Delta_t\left|\mathcal{I}\right|}$ 
		\Statex \;\;\;\;\;\;\;\;\;\;\;\;\textbf{endfor}
		\Statex \;\;\;\;\;\;\;\;\textbf{endfor}
		\Statex \;\;\;\; \textbf{end}
		\State \textbf{Calculate} $\hat{\mathbf{p}}_{0}:$
		\Statex \;\;\;\; \textbf{for} $k=1:N_K$
		\Statex \;\;\;\;\;\;\;\;\textbf{for} $i$ in $\mathcal{I}$
		\Statex \;\;\;\;\;\;\;\;\;\;\;\; $\displaystyle\hat{\mathbf{p}}_{0}=\hat{\mathbf{p}}_{0}+\frac{\hat{\mathbf{p}}_{i,k}-(k-1)\hat{\mathbf{v}}_u\Delta_t-(i-1)\frac{\lambda}{2}}{N_K\left|\mathcal{I}\right|}$ 
		\Statex \;\;\;\;\;\;\;\;\textbf{endfor}
		\Statex \;\;\;\;\textbf{endfor}
		\State \textbf{Calculate:} $\left\{\hat{\delta}_b,\hat{\epsilon}_b\right\}_{\forall b}$
		\Statex \;\;\;\; \textbf{for} $b=1:N_B$
		\Statex \;\;\;\;\;\;\;\;
		$\hat{\delta}_b=\frac{1}{N_U N_K}\sum_u \sum_k \left(\hat{\tau}_{bu,k}-\tau\left(\hat{\mathbf{p}}_0,\hat{\mathbf{v}}_u,\hat{\bar{\mathbf{s}}}\right)\right)$ 
		\Statex \;\;\;\;\;\;\;\;
		$\hat{\epsilon}_b=\frac{1}{N_U N_K}\sum_u \sum_k \left(\hat{f}_{d,bu,k}-f\left(\hat{\mathbf{p}}_0,\hat{\mathbf{v}}_u,\hat{\bar{\mathbf{s}}}\right)\right)$ 
		\Statex \;\;\;\;\textbf{endfor}
		\State \textbf{Output}:
		$\hat{\boldsymbol{\chi}}^{(0)}=\left[\hat{\mathbf{p}}_{U,0},\hat{\bar{\mathbf{v}}}_u,\left\{\hat{\delta}_b,\hat{\epsilon}_b\right\}_{b=1}^{N_B}\right]$, and $\hat{\bar{\mathbf{s}}}$.
	\end{algorithmic}
\end{algorithm}

\section{Numerical Results and Insights}
In this section, we present simulation results and discuss system design insights. The anchors were distributed uniformly in a sphere of radius $50{\rm m}$ around the receiver's central antenna element. We consider carrier frequencies in the range $f_c=1-30$ GHz and the antenna elements are separated by half a wavelength. The receiver speed was taken as $5{\rm ms^{-1}}$, and the transmitter velocity was selected as $10{\rm ms^{-1}}$. For the CRLB analysis, the bandwidth was chosen as $500$ MHz, and the operating SNR was maintained at $10$ dB. The number of anchors was chosen between $1$ and $5$, and the number of antenna elements was varied from $100$ to $400$. The time spacing between slots was varied from $0.2$s to $2$s. The transmitter emitted the following raised cosine pulse \cite{10476994}:
	\begin{equation}
		\begin{split}
			s(t)&=\frac{\cos\left(\pi \beta \frac{t}{T_s}\right)}{1-\left(2\beta\frac{t}{T_s}\right)^2} \frac{\sin\left(\pi \frac{t}{T_s}\right)}{\pi \frac{t}{T_s}}, \\
			S(f)&= \\
            &\begin{cases}
				T_s,&  0\leq |f| \leq \frac{1-\beta}{2T_s} \\
				\frac{T_s}{2}\left\{1+\cos\left[\frac{\pi T_s}{\beta}\left(|f|-\frac{1-\beta}{2T_s}\right)\right]\right\}, &\frac{1-\beta}{2T_s}\leq |f| \leq  \frac{1+\beta}{2T_s} \\
				0, & |f|>\frac{1+\beta}{2T_s}
			\end{cases},
		\end{split}
	\end{equation}where $T_s$ is the time for zero-crossings, and $\beta$ is related to bandwidth as:
\begin{equation}
	{\rm BW}=\frac{1+\beta}{T_s}.
\end{equation} The position (PEB), velocity (VEB), and orientation (OEB) error bounds were obtained using \eqnref{eqn:ccrb} as:
\begin{equation}
	\begin{split}
		{\rm PEB}&={\rm Trace}\left({\rm CCRB}(\boldsymbol{\kappa}_1)[1:3]\right), \\
		{\rm VEB}&={\rm Trace}\left({\rm CCRB}(\boldsymbol{\kappa}_1)[4:6]\right), \\
		{\rm OEB}&={\rm Trace}\left({\rm CCRB}(\boldsymbol{\kappa}_1)[7:9]\right).
	\end{split}
\end{equation}

\begin{figure*}[!htbp]
	\centering   
	\subfloat[]{\includegraphics[width=0.3333\linewidth,keepaspectratio]{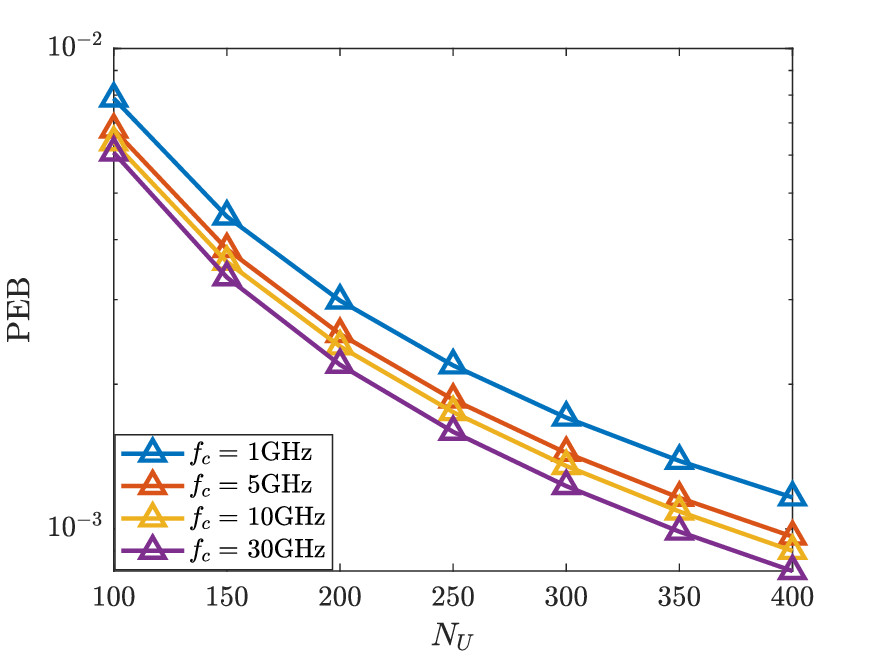}}
	\hfil
	\subfloat[]{\includegraphics[width=0.3333\linewidth,keepaspectratio]{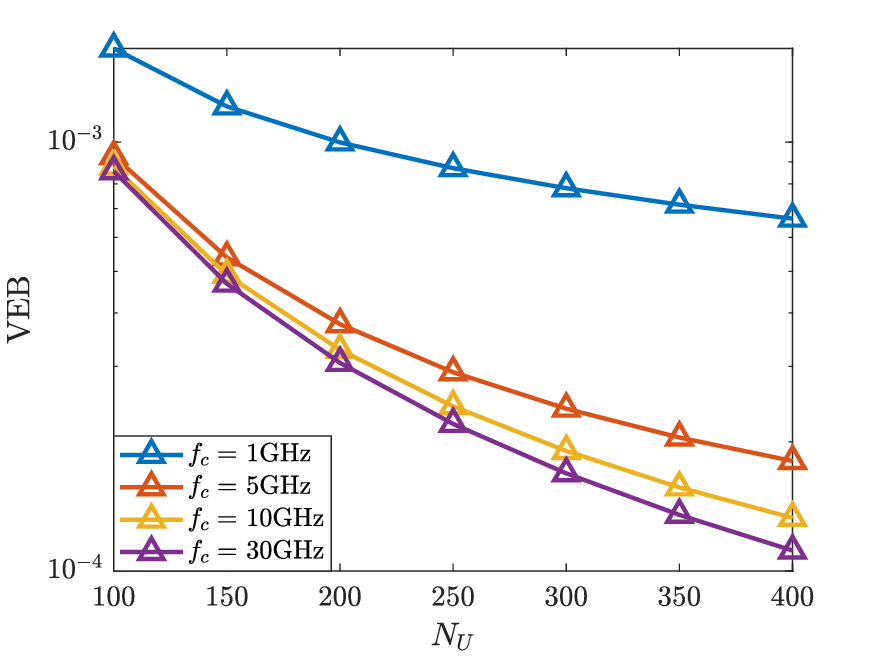}}
	\hfil
	\subfloat[]{\includegraphics[width=0.3333\linewidth,keepaspectratio]{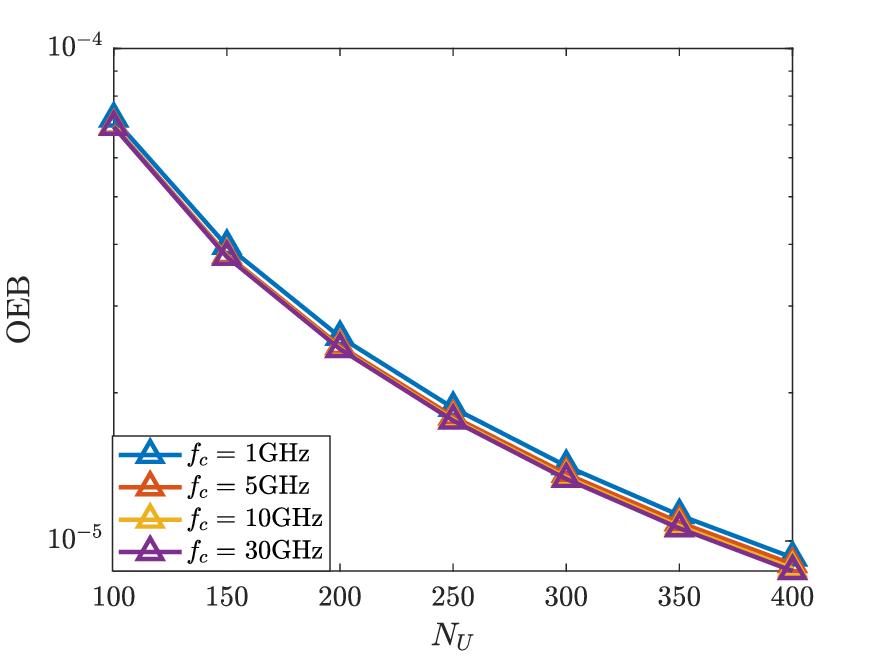}}
	\caption{C-CRB bounds with varying $N_U$ and $f_c$. $N_K=2$, $N_B=5$, and $\Delta_t=0.5 {\rm s}$.}
	\label{fig:vary_fc_2}
\end{figure*}
\begin{figure*}[!htbp]
	\centering   
	\subfloat[]{\includegraphics[width=0.3333\linewidth,keepaspectratio]{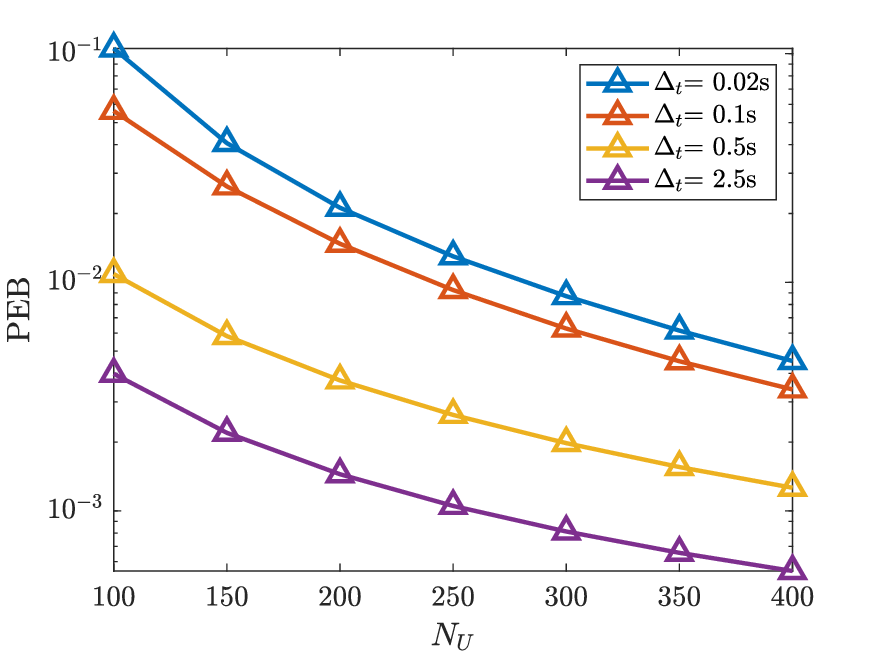}}
	\hfil
	\subfloat[]{\includegraphics[width=0.3333\linewidth,keepaspectratio]{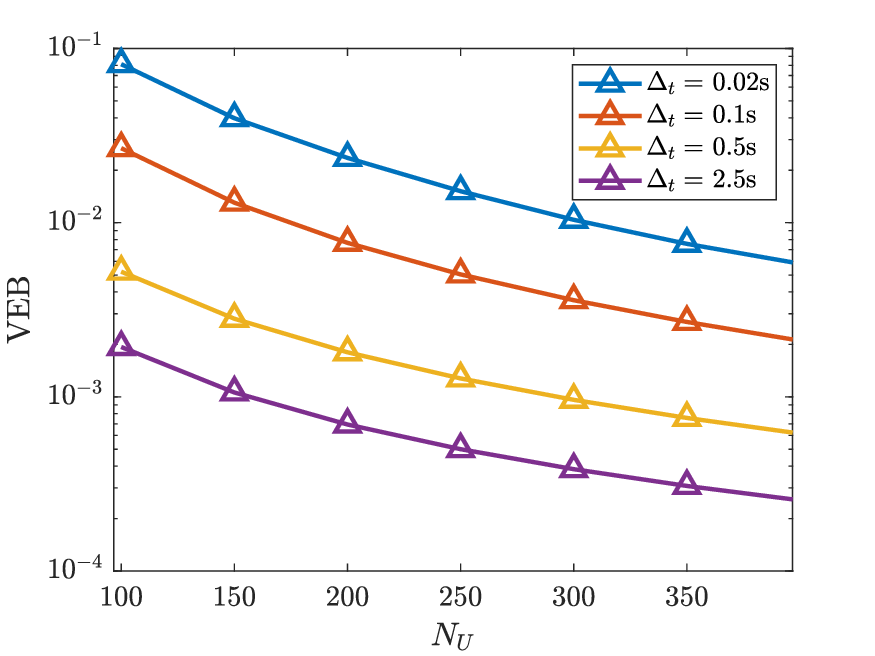}}
	\hfil
	\subfloat[]{\includegraphics[width=0.3333\linewidth,keepaspectratio]{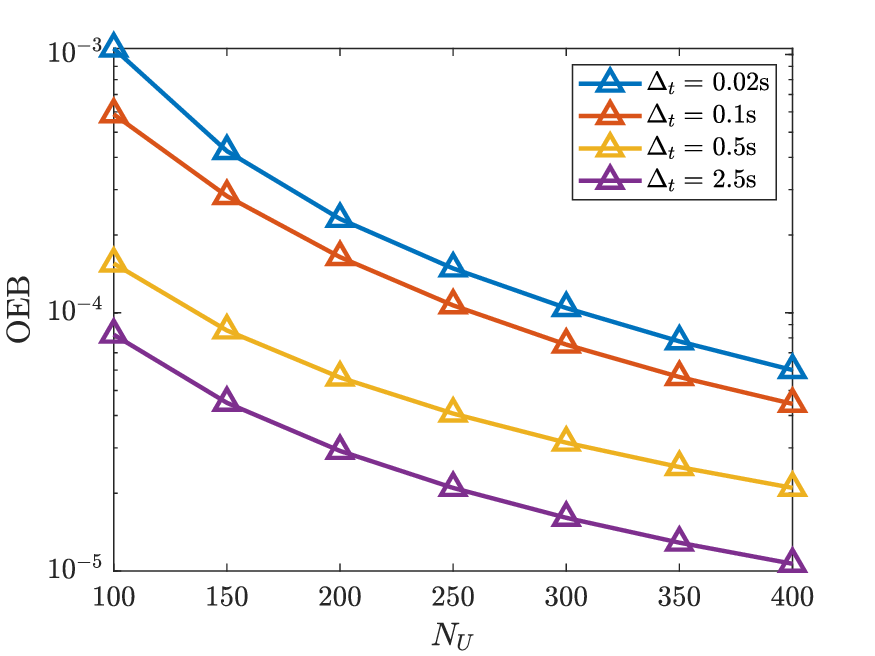}}
	\caption{C-CRB bounds with varying $N_U$ and $\Delta_t$. $f_c=10 {\rm GHz}$, $N_B=3$, and $N_K=2$.}
	\label{fig:vary_Dt}
\end{figure*}
\begin{figure*}[!htbp]
	\centering   
	\subfloat[]{\includegraphics[width=0.3333\linewidth,keepaspectratio]{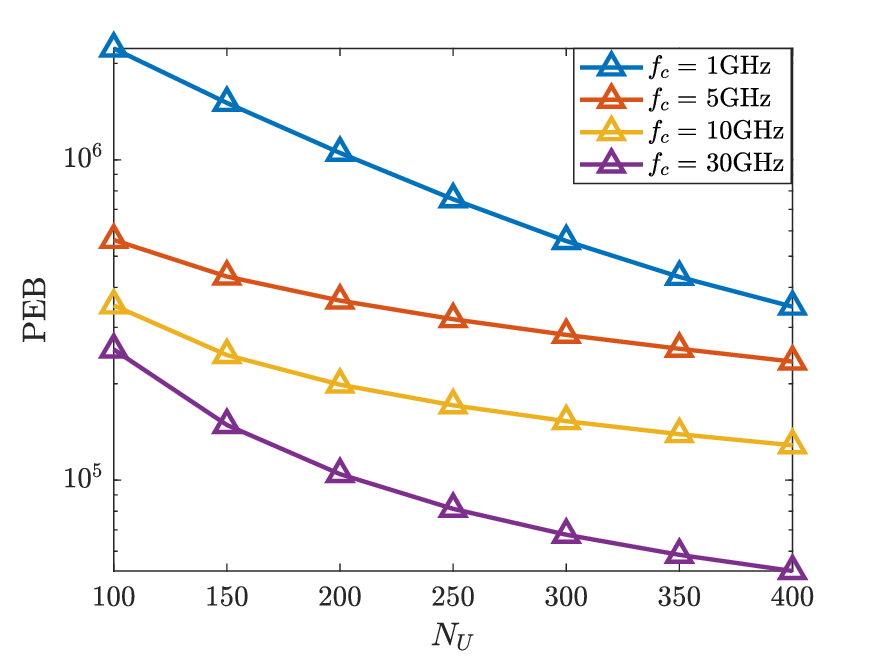}}
	\hfil
	\subfloat[]{\includegraphics[width=0.3333\linewidth,keepaspectratio]{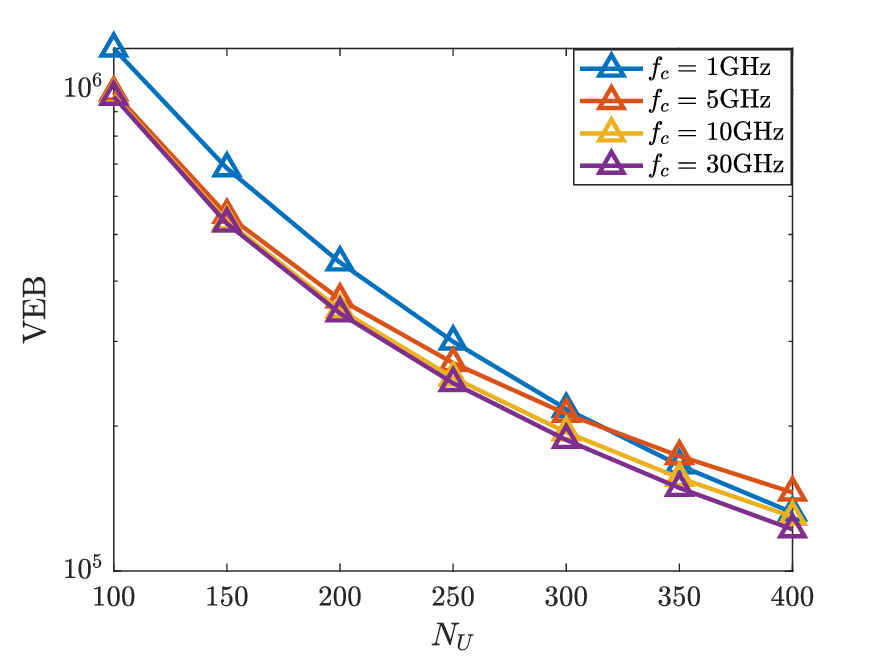}}
	\hfil
	\subfloat[]{\includegraphics[width=0.3333\linewidth,keepaspectratio]{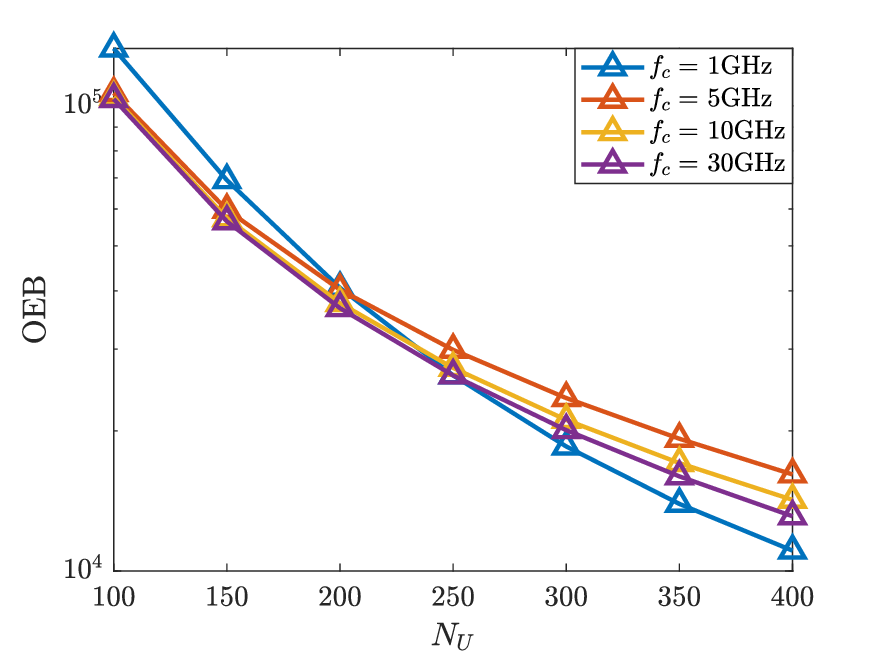}}
	\caption{C-CRB bounds with only Doppler measurements. $N_B=5$, $N_K=2$, and $\Delta_t=0.5$s.}
	\label{fig:dopp_only}
\end{figure*}
\begin{figure*}[!htbp]
	\centering   
	\subfloat[]{\includegraphics[width=0.3333\linewidth,keepaspectratio]{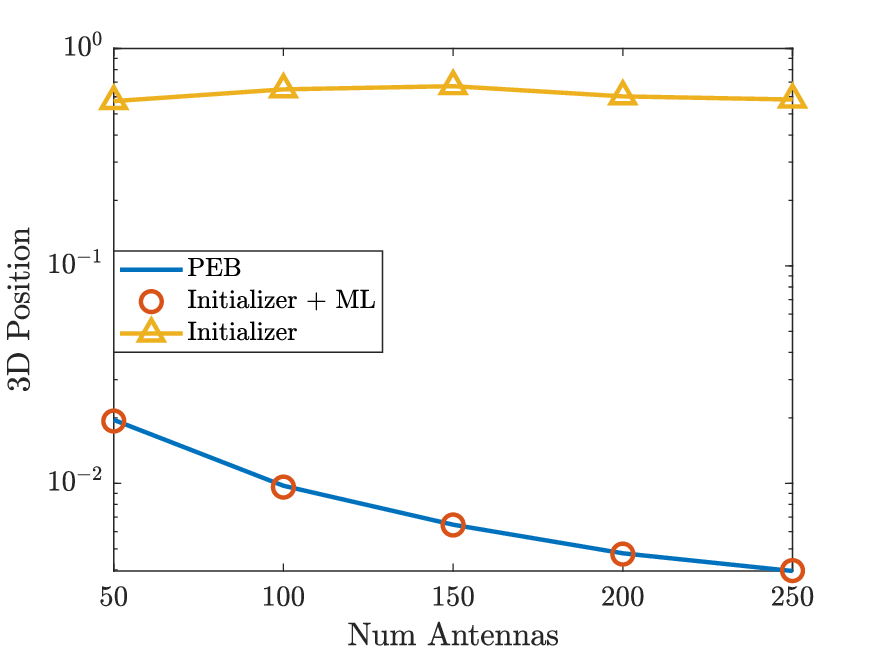}}
	\hfil
	\subfloat[]{\includegraphics[width=0.3333\linewidth,keepaspectratio]{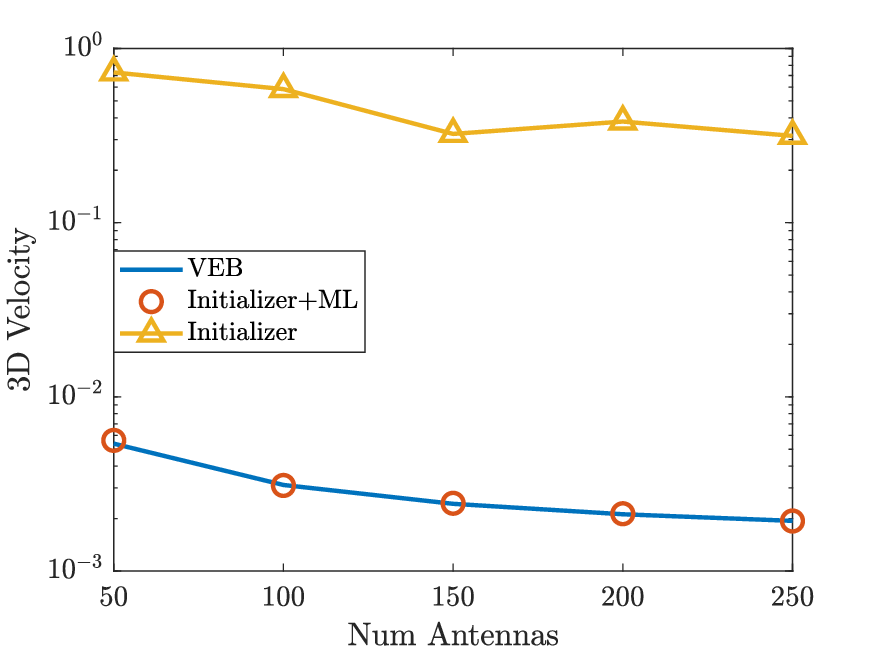}}
	\hfil
	\subfloat[]{\includegraphics[width=0.3333\linewidth,keepaspectratio]{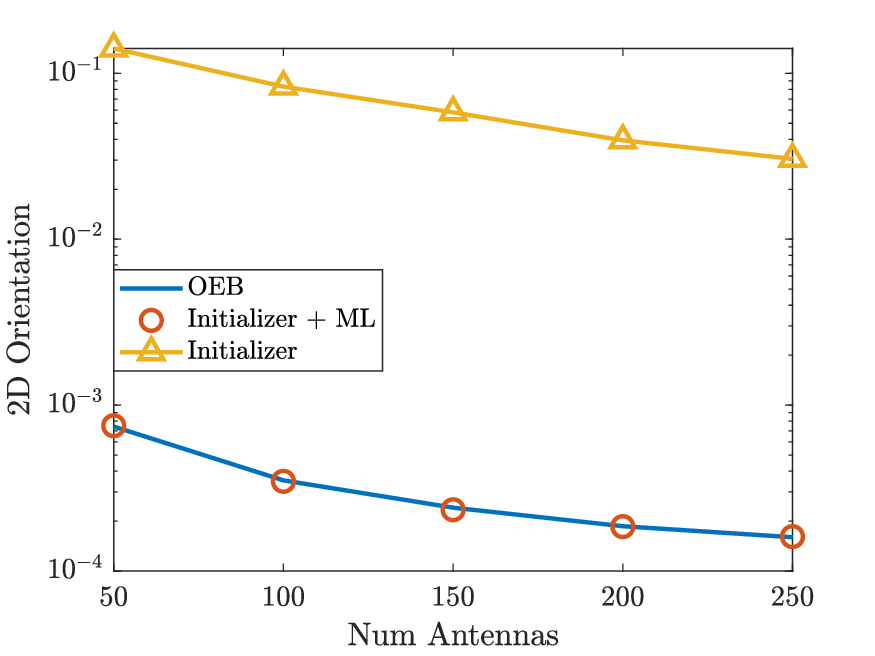}}
	\caption{ML estimator performance versus antenna elements. $f_c=1 {\rm GHz}$, $N_B=5$, $SNR=10$ dB, and $N_K=2$.}
	\label{fig:est_v_nu}
\end{figure*}
\begin{figure*}[!htbp]
	\centering   
	\subfloat[]{\includegraphics[width=0.3333\linewidth,keepaspectratio]{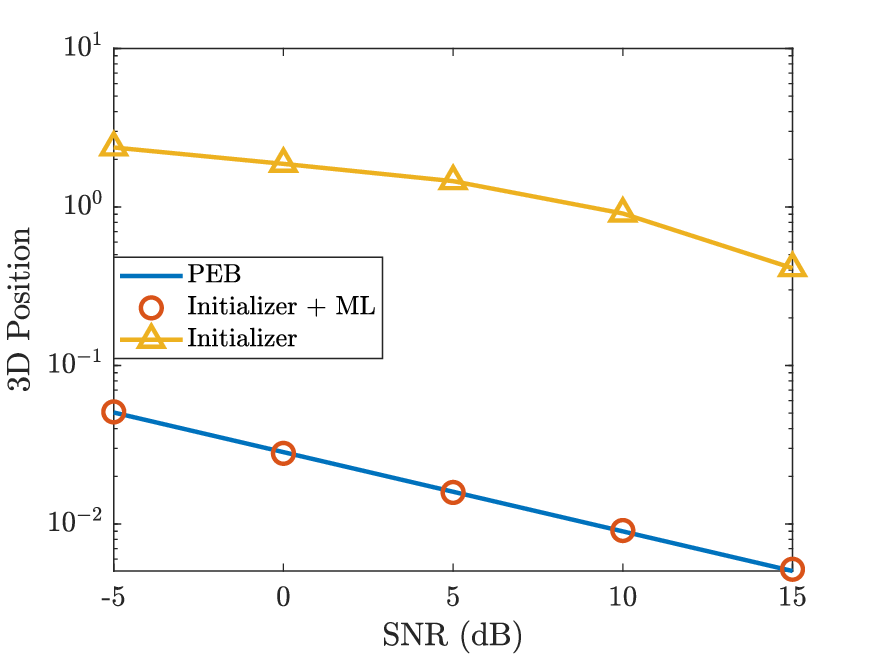}}
	\hfil
	\subfloat[]{\includegraphics[width=0.3333\linewidth,keepaspectratio]{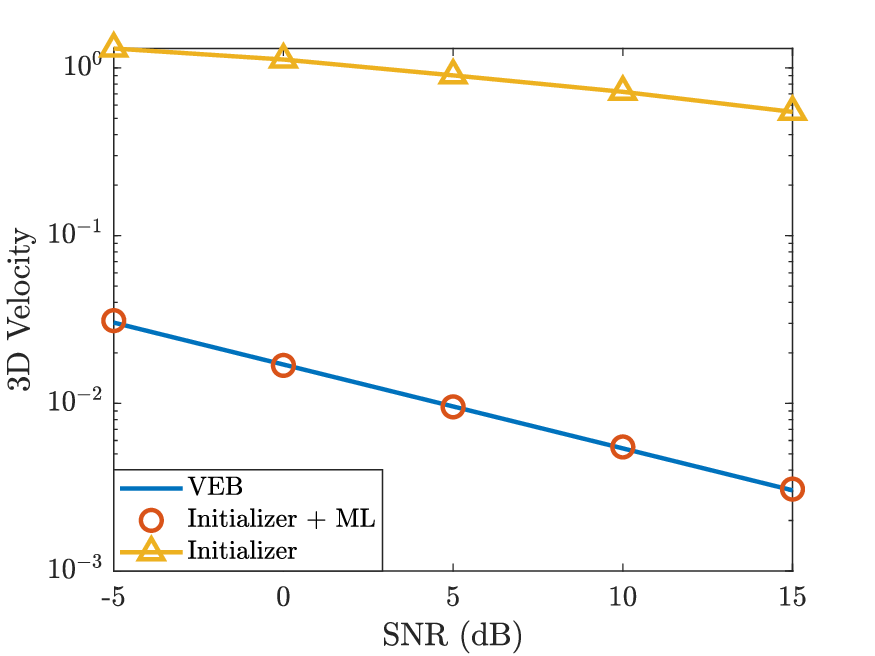}}
	\hfil
	\subfloat[]{\includegraphics[width=0.3333\linewidth,keepaspectratio]{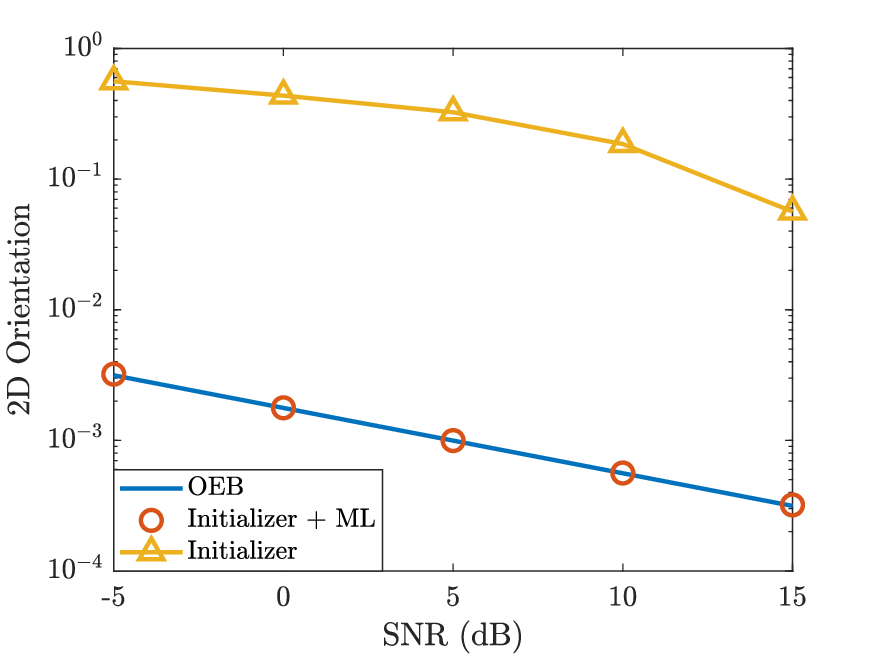}}
	\caption{ML estimator performance versus SNR. $f_c=1 {\rm GHz}$, $N_B=5$, $N_U=50$, and $N_K=2$.}
	\label{fig:est_v_snr}
\end{figure*}
\begin{figure*}[!htbp]
	\centering   
	\subfloat[]{\includegraphics[width=0.3333\linewidth,keepaspectratio]{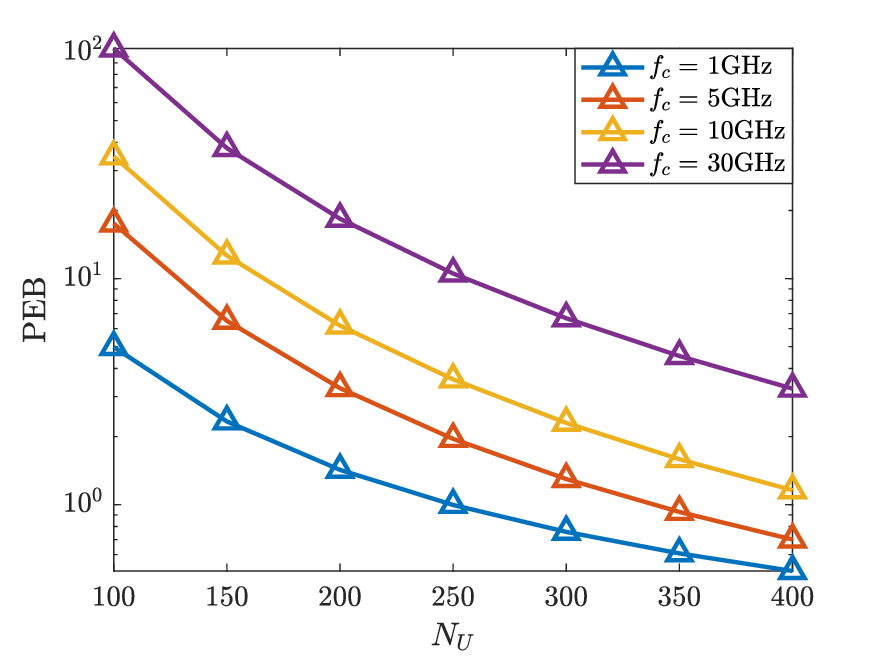}}
	\hfil
	\subfloat[]{\includegraphics[width=0.3333\linewidth,keepaspectratio]{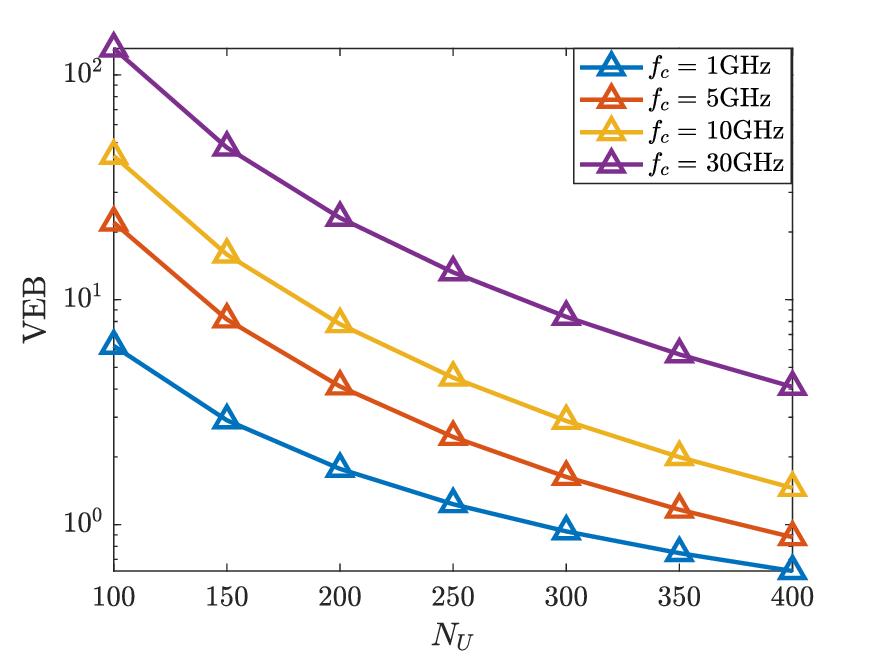}}
	\hfil
	\subfloat[]{\includegraphics[width=0.3333\linewidth,keepaspectratio]{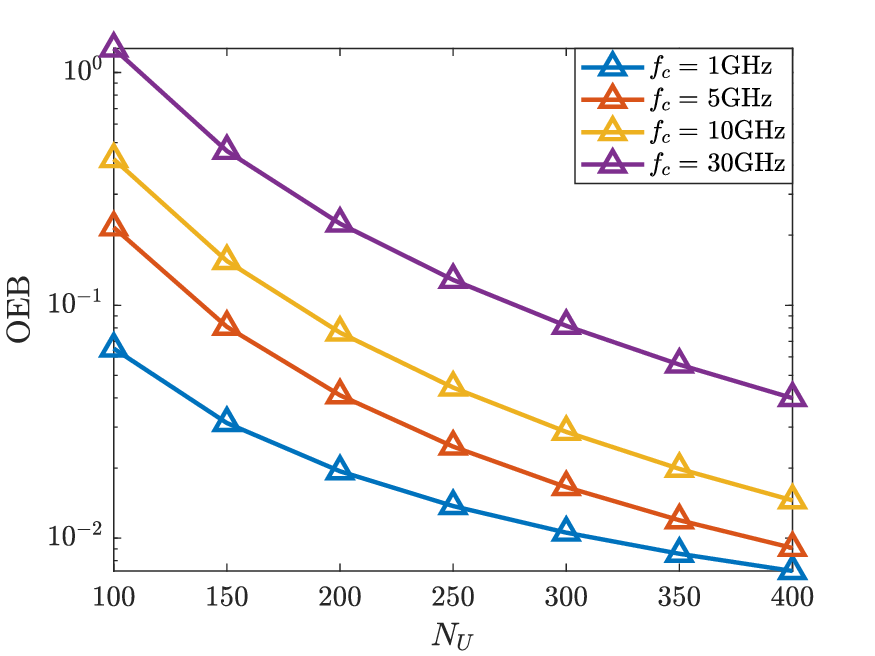}}
	\caption{Single anchor localization performance limits: $N_U$ and $f_c$. $N_B=1$, $N_k=4$ and $\Delta_t=0.5$s.}
	\label{fig:single_anchor}
\end{figure*}


The key takeaways from simulation studies are as follows:
\begin{enumerate}
	\item C-CRB analysis insights:
	\begin{enumerate}[label=(\alph*)]
		\item The impact of different carrier frequencies on error bounds depends on the anchor geometry. Generally, we see that for $N_B\geq5$, higher frequencies enable more precise localization, as illustrated in \figref{fig:vary_fc_2}. For $N_B\leq4$, the observation is not straightforward, and the error bounds do not show a monotonic trend with carrier frequencies. At a constant operating SNR, the Fisher information in delay measurements grows with the square of Doppler frequency. Increasing $f_c$ also enhances Doppler sensitivity. With well conditioned geometry which is enabled by a high number of anchors ($N_B\ge5$), these effect dominates and the bounds tighten monotonically. For smaller $N_B$, information loss due to nuisance parameters, especially channel gain, starts dominating which yields inconclusive trends. 
		
		
		\item In ~\figref{fig:vary_Dt}, we evaluate the effect of inter-slot spacing. Larger values of $\Delta_t$ leads to improved geometric diversity because the anchor positions are significantly different between time slots.  This produces diverse variations in observed delays, and the element wise angle of incidence observed at the different ELAA elements. This improves the conditioning of the transformation matrix $\mathbf{J}_{\boldsymbol{\eta}\rightarrow\boldsymbol{\kappa}}$, which improves the localization performance. 
		The effect of the spacing between time slots is more pronounced for velocity sensing as evident from Appendix~\ref{appn:A}. In general, increasing inter-slot spacing too much, can worsen the positioning performance as the propagation increasingly resembles the FF regime.
		
		\item  The addition of an each extra anchor improves geometric diversity and contributes an additive term to the FIM. Similarly, more snapshots provide additional measurements and contribute additively to the FIM. Thus an improved positioning performance is
        observed with more anchors/snapshots.
		\item Increasing antenna elements in an ELAA provides an improvement in localization performance due to an increase in the number of available channel parameters. In general, a larger aperture enables better localization by extending the Fraunhofer distance for a given carrier frequency. 
	\end{enumerate}
	\item Impact of Doppler measurements: In ~\figref{fig:dopp_only}, we analyze the exclusive role of Doppler measurements for full-motion state localization in the presence of unknown channel gains and frequency offsets. Our investigation reveals that despite offering theoretical estimation capability, Doppler measurements alone do not suffice to yield a reasonable localization accuracy. To establish our findings amongst existing research efforts around Doppler based NF velocity sensing \cite{perf_bound_vel_est_elaa}, we would like to apprise the readers that our model significantly differs in the following ways: we do not assume that delay (or range) and AoA has already been estimated and we have considered information loss due to unknown channel gains and frequency offsets. The first assumption is critical since it provides rich information to establish the LoS unit direction vector from the anchors to ELAA elements. We conclude that despite offering newer theoretical estimation capabilities, Doppler measurements contribute little to full-motion state information, and for the best operation, the use of other measurements, such as delay and AoA, is crucial.
	\item $8$D Localization algorithm: 
	\begin{enumerate}[label=(\alph*)]
		\item The variance for the delay $\sigma_\tau$ and Doppler frequency $\sigma_d$ estimator was taken as the CRLB obtained from \eqnref{eqn:fim_entry}.
		
		\item In the initialization step, to counter the effect of the delay offsets, we transformed the ToA measurements $\left(\{\hat{\delta}_{bi,k}\}_{\forall b,k},\, i\in\mathcal{I}\right)$ to time difference of arrival (TDoA) measurements and used a linear least-squares (L-LS) approach \cite{handbook_of_loc} to compute the position of the respective antenna elements. Our approach requires at least four anchors to arrive at a non-trivial solution.
		
		\item The plots of the estimator performance against $N_U$ and SNR are shown in \figref{fig:est_v_nu} and \figref{fig:est_v_snr}, respectively. The proposed estimator shows optimal convergence to the C-CRB limits.
	\end{enumerate}
	
	\item Minimum Infrastructure Insights:
	\begin{enumerate}[label=(\alph*)]
		\item Invertibility analysis of the EFIM reveals that at least three anchors are required for single-snapshot localization. With two snapshots, localization can be achieved with two anchors. Finally, single-anchor localization is not achievable for the considered system configuration. This is because the array sees the anchor location over different snapshots as collinear, leading to poor geometric dilution of precision (GDOP). However, if the anchors change their direction of motion in every snapshot then single anchor localization can be done with four or more snapshots. Single anchor performance limits are illustrated in ~\figref{fig:single_anchor}.
		\item It is possible to estimate the velocity with a single-snapshot setup. This is possible due to the availability of Doppler measurements. We have the following two key takeaways which can also be verified from Appendix~\ref{appn:A}:
        \begin{enumerate}
            \item It was previously impossible to perform single-snapshot velocity estimation with delay measurements alone since the velocity sensitivity is zero for $k=1$.
            \item For $k=1$, velocity estimation depends solely on the Doppler measurements. As discussed earlier, due to the limited information content of Doppler measurements, for the single-snapshot setup velocity estimation capabilities will only be coarse. Whereas, since position and orientation have finite delay sensitivity, they can be estimated finely.
        \end{enumerate}
	\end{enumerate}
\end{enumerate}

\section{Conclusions}
In this work, we demonstrated the potential of ELAAs to achieve high accuracy full motion state localization by leveraging the parameter rich NF operation. This capability opens the door to new applications and extended link ranges that were previously constrained by the limited observability of FF models. We derived the spherical-wave Doppler model for the NF and demonstrated that its spatial variations across the ELAA help recover all three velocity components. We analyzed the fundamental performance limits through a FIM analysis and investigated the effects of different operating parameters, such as the number of antenna elements, number of anchors, operating frequency, and inter-slot spacing, to develop system-level insights. We demonstrated that the minimum infrastructure required to achieve full motion state recovery consists of at least two anchors with two snapshots, or three anchors with a single snapshot, or a single anchor with four snapshots and different directions of anchor velocity in each snapshot for the single snapshot case. We also demonstrated the possibility of single-snapshot velocity estimation, which was previously not possible without Doppler measurements in the NF. Furthermore, we revealed that standalone Doppler measurements do not suffice to enable rich NF localization because they cannot overcome the information losses due to unknown channel gains and frequency offsets, and for ideal operation another information rich set of measurements, such as delay or AoA is required. 

We developed a ML based approach for the joint estimation of motion state parameters and synchronization offsets from measured channel characteristics. To solve the non-convex ML estimation problem, we developed a geometric initializer that leverages the array's linear geometry and the system's rectilinear kinematics to provide good initial guesses for the ML estimation module. Our estimator is shown to achieve optimal convergence to the C-CRB limits.

Promising directions for future work include moving beyond Gaussian noise assumption by developing robust estimators and analyzing corresponding performance bounds under impulsive or heavy-tailed noise scenarios. Another extension is NF sensing with unknown or partially known transmit signals \cite{8552385}, which calls for blind or semi-blind delay–Doppler processing and joint estimation of the motion parameters and transmitted data.

\appendices
\section{Jacobian Construction}
\label{appn:A}
We derive the respective gradients used in the line search routine, and for constructing the transformation Jacobian matrix $\mathbf{J}_{\boldsymbol{\eta}\rightarrow \boldsymbol{\kappa}}$. 

The derivatives with respect to delay are given as:
\begin{equation}
	\begin{split}
		&\nabla_{\mathbf{p}_{U,0}}\tau_{bu,k} =\frac{-\boldsymbol{\Delta}_{bu,k}}{c}, \\
		&\nabla_{\mathbf{\bar{v}}_U} \tau_{bu,k}, 
		= \frac{-(k-1)\Delta_t }{c} \boldsymbol{\Delta}_{bu,k}, \\
		&\nabla_{\bar{\mathbf{s}}} \tau_{bu,k} 
		= \frac{-(u-1)\lambda}{2c}\boldsymbol{\Delta}_{bu,k}, \\
		&\nabla_{\delta_b}\tau_{bu,k} =1, \\
		&\nabla_{\epsilon_b}\tau_{bu,k} =0.
	\end{split}
\end{equation}

The derivative with respect to Doppler shift are given as: 
\begin{equation}
	\begin{split}
		&\nabla_{\mathbf{p}_{U,0}}f_{d,bu,k} =\frac{f_c}{c}\frac{1}{d_{bu,k}}\left[\mathbf{I}_3-\boldsymbol{\Delta}_{bu,k}\boldsymbol{\Delta}_{bu,k}^{\rm T}\right]\bar{\mathbf{v}}^o_{b,k}, \\
		&\nabla_{\mathbf{\bar{v}}_U} f_{d,bu,k}, 
		=\frac{f_c}{c}\Bigg[\boldsymbol{\Delta}_{bu,k}\Bigg.\\
		&\left.+\frac{(k-1)\Delta_t\left(\mathbf{I}_3-\boldsymbol{\Delta}_{bu,k}\boldsymbol{\Delta}_{bu,k}^{\rm T}\right)\bar{\mathbf{v}}^o_{b,k}}{d_{bu,k}}\right], \\
		&\nabla_{\bar{\mathbf{s}}} f_{d,bu,k} 
		= \frac{f_c(u-1)\lambda}{2cd_{bu,k}}\left[\mathbf{I}_3-\boldsymbol{\Delta}_{bu,k}\boldsymbol{\Delta}_{bu,k}^{\rm T}\right]\bar{\mathbf{v}}^o_{b,k}, \\
		&\nabla_{\delta_b}f_{d,bu,k} =0, \\
		&\nabla_{\epsilon_b}f_{d,bu,k} =1.
	\end{split}
\end{equation}

The derivative with respect to channel gain are given as:
\begin{equation}
	\begin{split}
		&\nabla_{\mathbf{p}_{U,0}}\beta_{bu,k} =\frac{\alpha \lambda d_{bu,k}^{-\alpha-1}\boldsymbol{\Delta}_{bu,k}}{4\pi}, \\
		&\nabla_{\mathbf{\bar{v}}_U} \beta_{bu,k}, 
		=\frac{\alpha \lambda d_{bu,k}^{-\alpha-1}\left(k-1\right)\Delta_t\boldsymbol{\Delta}_{bu,k}}{4\pi}, \\
		&\nabla_{\bar{\mathbf{s}}} \beta_{bu,k} 
		= \frac{\alpha\lambda}{4\pi}d_{bu,k}^{-\alpha-1}\boldsymbol{\Delta}_{bu,k}(u-1)\frac{\lambda}{2},\\
		&\nabla_{\delta_b}\beta_{bu,k} =0, \\
		&\nabla_{\epsilon_b}\beta_{bu,k} =0.
	\end{split}
\end{equation}

\bibliographystyle{IEEEtran}
\bibliography{bib}
\vspace{12pt}
\end{document}